\documentclass[aps,prd,preprint,groupedaddress,preprintnumbers,longbibliography,nofootinbib]{revtex4-1}

\newcommand{\vvev}[1]{\left\langle\kern-0.3em\left\langle #1
    \right\rangle\kern-0.3em\right\rangle}

\usepackage{amsmath}
\usepackage{graphicx}
\usepackage{amsfonts}
\usepackage{caption}
\usepackage{subcaption}
\usepackage{xcolor}

\usepackage{cancel}

\usepackage{mathrsfs}

\usepackage[cal=esstix,frak=euler,scr=boondox,bb= pazo]{mathalfa}
\newcommand{\xcal}{\mathcal{x}}
\newcommand{\ycal}{\mathcal{y}}
\newcommand{\xs}{X}
\newcommand{\ys}{Y}

\begin{document}


\title{Time evolution of density matrices as a theory of random surfaces}

\author{Carlo Pagani}
\email[]{cpagani@uni-mainz.de}
\affiliation{Institut f\"{u}r Physik (THEP),
  Johannes-Gutenberg-Universit\"{a}t\\ Staudingerweg 7, 55099 Mainz,
  Germany}
%
\author{Martin Reuter}
\email[]{reutma00@uni-mainz.de}
\affiliation{Institut f\"{u}r Physik (THEP),
  Johannes-Gutenberg-Universit\"{a}t\\ Staudingerweg 7, 55099 Mainz,
  Germany}



\begin{abstract}
In the operatorial formulation of quantum statistics, the time evolution
of density matrices is governed by von Neumann's equation. Within
the phase space formulation of quantum mechanics it translates into
Moyal's equation, and a formal solution of the latter is provided
by Marinov's path integral. In this paper we uncover a hidden property
of the Marinov path integral, demonstrating that it describes a theory
of ruled random surfaces in phase space.
\end{abstract}


\maketitle

\section{Introduction}

A quantum mechanical system can be conviniently described by a density matrix. Besides its conceptual relevance, the density matrix plays a key role in modern applications of quantum mechanics, 
such as quantum computing \cite{nielsen_chuang_2010} and the study of open quantum systems \cite{Breuer:2002pc}. 
A powerful approach to study the time evolution of density matrices is provided by path integral methods. 
This requires to extend the Feynman path integral, which provides the evolution kernel for pure states, by adopting a suitable formalism such as the Schwinger-Keldysh one \cite{Schwinger:1960qe,Keldysh:1964ud}.

An object closely related to the density matrix is the Wigner function, whose time evolution is governed by the so-called Moyal equation \citep{Wigner:1932eb,GROENEWOLD1946405,Moyal:1949sk}. 
Associated with the Moyal equation there exists a formal solution provided by a functional integral \cite{MARINOV1991}, the Marinov integral. 
This functional approach has been recently revisted in \cite{Gozzi:2020wef} where it has been shown that it allows one to construct novel approximations \cite{Gozzi:2020wef}. 
Among other things, it turns out that the complex instantons appearing in standard Feynman path integrals \cite{Dunne:2015eaa,Dunne:2016nmc,Dunne:2016qix} play a role in Marinov integral too.

In this work we put forward a novel formalism in which the evolution kernel of Wigner functions is provided by a functional integral over random surfaces, 
rather than random trajectories in phase space. 
The formulation of Marinov's path integral for a quantum system consisting of $N$ degrees of freedom requires the functional integration over $2N$ phase space variables plus further $2N$ conjugate variables. An appealing feature of the formalism constructed in this paper is that the $4N$ variables characterizing Marinov's integral can be embedded into $2N$ functions defining a surface, as opposed to functions of time alone.

The paper is organized as follows. 
In section \ref{sec:Marinov_PI} we set up our notations and conventions and review Marinov's integral. 
In section \ref{sec:rnd_surface_representation} we derive the random surface representation of the evolution kernel of Wigner functions. 
In section \ref{sec:sum_classical_surf} we apply our formalism in the semiclassical limit and discuss how quantum phenomena, such as interference, arise in this context. 
We summarize our findings in section \ref{sec:conclusions}.
Some technical details are confined in an Appendix.

\section{The Marinov path integral} \label{sec:Marinov_PI}

In this section we collect a number of preliminaries from classical
and quantum statistical mechanics before introducing Marinov's path
integral for the time evolution of Wigner functions.

\subsection{Mechanics {\`a} la Hamilton}

A generic Hamiltonian system with $N$ degrees of freedom is defined
by a triple $\left({\cal M},\omega,H\right)$ where ${\cal M}$ is
a $2N$-dimensional smooth manifold, $\omega\equiv\frac{1}{2}\omega_{ab}d\phi^{a}\wedge d\phi^{b}$
a closed non-degenerate two-form, and $H$ a scalar function on ${\cal M}$.
These ingredients are commonly referred to as the system's phase space,
symplectic two-form, and Hamiltonian function, respectively \cite{AbrahamMarsden_CM}. 

In the following we focus on the simple case where ${\cal M}$ is
the cotangent bundle of a $N$-dimensional configuration manifold,
and we assume the latter to be $\mathbb{R}^{N}$ so that we may identify
${\cal M}$ with $\mathbb{R}^{N}\times\mathbb{R}^{N}$. In generic
(local) coordinates on ${\cal M}$, denoted $\phi^{a}\,,\,a=1,\cdots,2N$,
the two-form components $\omega_{ab}\left(\phi\right)$ have a non-trivial
$\phi$-dependence in general. In the case at hand, however, it is
possible to choose Darboux coordinates on all of ${\cal M}$; the $\omega_{ab}$
are constants then, assuming values $0$ or $\pm1$ only. 

Henceforth we assume that the coordinates $\phi^{a}=\left(p^{1},\cdots,p^{N},q^{1},\cdots,q^{N}\right)$
are chosen such that the $2N\times2N$ matrix ($\omega_{ab}$) consists
of the $N\times N$ blocks $\omega_{pq}=-\omega_{qp}=\mathbb{I}$,
$\omega_{pp}=\omega_{qq}=0$. In this basis the inverse matrix $\left(\omega^{ab}\right)=\left(\omega_{ab}\right)^{-1}$,
providing the components of the Poisson tensor, displays the blocks
$\omega^{qp}=-\omega^{pq}=\mathbb{I}$, $\omega^{pp}=\omega^{qq}=0$.
In this notation, the Poisson bracket of two functions on ${\cal M}$
is given by $\left\{ f_{1},f_{2}\right\} _{{\rm P}}=\partial_{a}f_{1}\omega^{ab}\partial_{b}f_{2}$,
the components of the Hamiltonian vector field $h=h^{a}\partial_{a}$,
$\partial_{a}\equiv\frac{\partial}{\partial\phi^{a}}$, read $h^{a}\left(\phi\right)=\omega^{ab}\partial_{b}H\left(\phi\right)$,
and Hamilton's equation of motion assumes the form $\dot{\phi}^{a}\left(t\right)=h^{a}\left(\phi\left(t\right)\right)$
\cite{AbrahamMarsden_CM}.

\subsection{Quantum mechanics {\`a} la Schr{\"o}dinger}

Let us consider an arbitrary quantum system with $N$ degrees of freedom
whose classical limit is described by a Hamiltonian system of the type
above. The quantum mechanical time evolution is governed by a certain
(time-independent) Hamiltonian operator $\hat{H}$. On the states
$|\psi\rangle$ of the corresponding Hilbert space it acts via $|\psi\left(T\right)\rangle=\widehat{{\cal U}}\left(T;T_{0}\right)|\psi\left(T_{0}\right)\rangle$
whereby the evolution operator $\widehat{{\cal U}}\left(T;T_{0}\right)$
satisfies Schr{\"o}dinger's equation $i\hbar\partial_{T}\widehat{{\cal U}}\left(T;T_{0}\right)=\widehat{H}\widehat{{\cal U}}\left(T;T_{0}\right)$,
with $\widehat{{\cal U}}\left(T_{0};T_{0}\right)=1$. The associated
integral kernel for position space, or ``Feynman kernel'', writes
\begin{eqnarray*}
K\left(x^{\prime},T;x^{\prime\prime},T_{0}\right) & = & \langle x^{\prime}|\,\widehat{{\cal U}}\left(T;T_{0}\right)|x^{\prime\prime}\rangle
\end{eqnarray*}
where $x^{\prime},x^{\prime\prime}\in\mathbb{R}^{N}$ denote the eigenvalues
of the position operator $\widehat{q}$. (Here and in the following
we suppress configuration space indices.)

In this paper we describe pure and mixed states alike in terms of
their density operator $\widehat{\rho}\left(t\right)$. It satisfies
von Neumann's equation
\begin{eqnarray}
i\hbar\partial_{t}\widehat{\rho}\left(T\right) & = & -\left[\widehat{\rho}\left(T\right),\widehat{H}\right]\,,\label{eq:2-1_vonNeumann-evol-density-mat}
\end{eqnarray}
which is formally solved by
\begin{eqnarray}
\widehat{\rho}\left(T\right) & = & \widehat{{\cal U}}\left(T;T_{0}\right)\widehat{\rho}\left(T_{0}\right)\widehat{{\cal U}}\left(T;T_{0}\right)^{\dagger}\,,\label{eq:2-2-density-mat_via_evol-operator}
\end{eqnarray}
or more explicitly, in terms of matrix elements, by:
\begin{eqnarray}
\langle x^{\prime}|\widehat{\rho}\left(T\right)|x^{\prime\prime}\rangle & = & \int d^{N}y^{\prime}d^{N}y^{\prime\prime}K\left(x^{\prime},T;y^{\prime},T_{0}\right)K\left(x^{\prime\prime},T;y^{\prime\prime},T_{0}\right)^{*}\langle y^{\prime}|\widehat{\rho}\left(T_{0}\right)|y^{\prime\prime}\rangle\,.\label{eq:2-12}
\end{eqnarray}

\subsection{Evolving pure states {\`a} la Feynman}

The above time evolution kernel $K$ possesses a well known representation
in terms of a Feynman-type functional integral. Allowing for an arbitrary
Hamiltonian operator $\widehat{H}$ which is not necessarily quadratic
in the momenta, the integration extends over paths in phase space
\cite{Schulman_pi_book}:
\begin{eqnarray}
K\left(x^{\prime},T;x^{\prime\prime},T_{0}\right) & = & \int_{x\left(T_{0}\right)=x^{\prime\prime}}^{x\left(T\right)=x^{\prime}}{\cal D}x\left(\cdot\right)\int{\cal D}p\left(\cdot\right)\exp\left(\frac{i}{\hbar}\int_{T_{0}}^{T}dt\left\{ p\left(t\right)\dot{x}\left(t\right)-H\left(p\left(t\right),x\left(t\right)\right)\right\} \right)\,.\label{eq:2-13-phase-space-PI}
\end{eqnarray}
This formal expression must be seen as the continuum limit of an appropriate
discrete multiple integral over a time lattice. Contrary to ordinary
integrals, path integrals in general do depend on the discretization
scheme underlying their definition. The crucial relationship between
this discretization scheme, the $c$-number function $H\left(p,x\right)$
under the path integral, and the Hamiltonian $\widehat{H}$ appearing
in the operator $\widehat{{\cal U}}\left(T;T_{0}\right)=\exp\left(-\frac{i}{\hbar}\widehat{H}\left(T-T_{0}\right)\right)$
whose matrix element is evaluated, can be summerized as follows 
\cite{Chaichian:2001cz}.

\noindent\textbf{(i)} Assume we are given an externally prescribed
operator $\widehat{H}$, and we apply the standard techniques for
passing from $\widehat{{\cal U}}\left(T;T_{0}\right)$ to the path
integral representation of the corresponding kernel $K\left(x^{\prime\prime},T;x^{\prime},T_{0}\right)$.
They involve discretizing the time axis and result in a sequence of
ordinary $n$-fold integrals that converges to the sought-for kernel
in the limit $n\rightarrow\infty$. 

Many different discretization schemes ``${\rm DS}_{i}$'' can be
employed for this purpose. As for the integrands thus constructed,
\emph{different discretization schemes }${\rm DS}_{i}$\emph{ give
rise to different $c$-number functions $H_{i}\left(p,x\right)$ for
one and the same operator $\widehat{H}$ }in general. The functions
$H_{i}$ and $H_{j}$, $i\neq j$, differ by terms of order $O\left(\hbar\right)$.

\noindent\textbf{(ii)} Conversely, assume we are given a $c$-number
function $H\left(p,x\right)$ and try to promote it to an operator
by substituting non-commuting operators for its eigenvalues, $p\rightarrow\widehat{p}$,
$x\rightarrow\widehat{q}$, with $\left[\widehat{q}^{j},\widehat{p}^{k}\right]=i\hbar\delta^{jk}$.
This leads to a uniquely defined operator, denoted $\left[H\left(\widehat{p},\widehat{q}\right)\right]_{{\rm OS}}$,
provided an extra input is supplied, namely an operator ordering scheme
(${\rm OS}$) which specifies the order in which positions and momenta
are to be written when translating $x$-$p$-monomials to operators.

Coming back to the externally prescribed operator $\widehat{H}$,
there will exist many pairs of functions $H_{\alpha}\left(p,x\right)$
and associated ordering schemes ${\rm OS}_{\alpha}$ that can be used
to represent one and the same operator: $\widehat{H}=\left[H_{1}\left(\widehat{p},\widehat{q}\right)\right]_{{\rm OS}_{1}}=\left[H_{2}\left(\widehat{p},\widehat{q}\right)\right]_{{\rm OS}_{2}}=\cdots$.
The functions $H_{\alpha}$ and $H_{\beta}$, $\beta\neq\alpha$,
differ by terms of order $O\left(\hbar\right)$. 

\noindent\textbf{(iii)} The key fact is that the discretization and
the operator ordering schemes are in direct correspondence. When we
fix $\widehat{H}$, and furthermore opt for a specific ordering scheme,
${\rm OS}_{*}$ say, with $\widehat{H}=\left[H_{*}\left(\widehat{p},\widehat{q}\right)\right]_{{\rm OS}_{*}}$,
then there exists a discretization scheme ${\rm DS}_{*}$ associated
to ${\rm OS}_{*}$ such that the sequence of integrals, when equipped
with the $c$-number function $H_{*}\left(p,x\right)$, converges
to the kernel $K$ based upon the operator $\widehat{H}$. Generically,
any other ordering scheme different from ${\rm OS}_{*}$ will not
lead to this kernel when used along with $H_{*}$.

Typical examples of such ${\rm OS}$-${\rm DS}$ correspondences include
the mid-point prescription for the discretization on the time lattice,
which belongs to Weyl ordering, and likewise the pre-point discretization
and post-point discretization, respectively, belonging to the $pq$-,
and $qp$-orderings, respectively. Given the $c$-number monomial
$xp$, for example, these ordering schemes translate it to the operators
$\frac{1}{2}\left(\widehat{p}\widehat{q}+\widehat{q}\widehat{p}\right)$,
$\widehat{p}\widehat{q}$, and $\widehat{q}\widehat{p}$, respectively
\cite{Chaichian:2001cz,Gro_Stei}.

\subsection{Quantum mechanics {\`a} la Wigner-Weyl-Moyal}

Given a system with quantum mechanical Hilbert space ${\cal H}$ and
classical phase space ${\cal M}$, the idea behind the Wigner-Moyal
phase space formulation of quantum mechanics is to represent the operators
on ${\cal H}$ in terms of complex valued functions on ${\cal M}$.
One defines a one-to-one ``symbol map'' between the operators $\widehat{A},\widehat{B},\cdots$
and their ``symbols'' $A\left(p,q\right),B\left(p,q\right),\cdots$.
This map is given the status of an algebra homomorphism by introducing
a ``star product'' among symbols such that $A\left(p,q\right)*B\left(p,q\right)$
is the phase space counterpart of the operator product $\widehat{A}\widehat{B}$.

There exist many such symbol maps. Henceforth we mostly employ \emph{Weyl
symbols} which offer a number of advantages;\footnote{The only exception is in the appendix where we also employ the pre-point discretization scheme, i.e., the $\hat{p}\hat{q}$-ordering and its associated symbol.} 
for example, Weyl symbols of Hermitian operators are always real.
In terms of the operator's configuration space matrix elements $\langle x^{\prime}|\widehat{A}|x^{\prime\prime}\rangle$,
the Weyl symbol map can be presented as
\begin{eqnarray}
A\left(p,q\right) & = & \int d^{N}s\,\langle q+\frac{s}{2}|\widehat{A}|q-\frac{s}{2}\rangle e^{-\frac{i}{\hbar}sp}\label{eq:2-11_Weyl-symb-map}
\end{eqnarray}
and its inverse reads
\begin{eqnarray}
\langle x^{\prime}|\widehat{A}|x^{\prime\prime}\rangle & = & \int\frac{d^{N}p}{\left(2\pi\hbar\right)^{N}}\,A\left(p,\frac{x^{\prime}+x^{\prime\prime}}{2}\right)e^{\frac{i}{\hbar}p\left(x^{\prime}-x^{\prime\prime}\right)}\,.\label{eq:2-12_map_from_symb_to_mat-element}
\end{eqnarray}
As above, $\left(p^{1},p^{2},\cdots,p^{N},q^{1},q^{2},\cdots,q^{N}\right)\equiv\left(\phi^{a}\right)$
are Darboux coordinates on ${\cal M}=\mathbb{R}^{N}\times\mathbb{R}^{N}$
here, and scalar products like $sp\equiv\sum_{i=1}^{N}s^{i}p^{i}$
are always understood. 

Taking advantage of the Poisson tensor $\omega^{ab}$, the pertinent
star product can be rewritten compactly as
\begin{eqnarray}
\left(A*B\right)\left(\phi\right) & = & A\left(\phi\right)\exp\left(i\frac{\hbar}{2}\overleftarrow{\partial}_{a}\omega^{ab}\overrightarrow{\partial}_{b}\right)B\left(\phi\right)\,=\,A\left(\phi\right)B\left(\phi\right)+O\left(\hbar\right)\,.\label{eq:2-13_def_Weyl_star_product}
\end{eqnarray}
The image of the commutator $\left[\widehat{A},\widehat{B}\right]/i\hbar$
under the symbol map is the Moyal bracket, a quantum deformation of
the Poisson bracket \cite{Bayen:1977ha,Bayen:1977hb}:
\begin{eqnarray}
\left\{ A,B\right\} _{{\rm M}} & = & \frac{1}{i\hbar}\left(A*B-B*A\right)\,=\,{\rm symb}\left(\frac{1}{i\hbar}\left[\widehat{A},\widehat{B}\right]\right)\label{eq:2-14_def_Moyal_bracket}\\
 & = & A\left(\phi\right)\frac{2}{\hbar}\sin\left(i\frac{\hbar}{2}\overleftarrow{\partial}_{a}\omega^{ab}\overrightarrow{\partial}_{b}\right)B\left(\phi\right)\,=\,\left\{ A\left(\phi\right),B\left(\phi\right)\right\} _{{\rm P}}+O\left(\hbar^{2}\right)\,.\label{eq:2-15_Moyal_bracket_via_exp-sin}
\end{eqnarray}

If we denote the symbols of some density operator $\widehat{\rho}$
and the Hamiltonian $\widehat{H}$ by $\rho\left(\phi\right)$ and
$H\left(\phi\right)$, respectively, von Neumann's equation (\ref{eq:2-1_vonNeumann-evol-density-mat})
translates to the \emph{Moyal equation}
\begin{eqnarray}
\partial_{T}\rho\left(\phi,T\right) & = & -\left\{ \rho,H\right\} _{{\rm M}}\,,\label{eq:2-19-Moyal-eq}
\end{eqnarray}
a quantum deformation of Liouville's equation used in classical statistical
mechanics.

In the special case of a pure quantum state $\widehat{\rho}=|\psi\rangle\langle\psi|$
the corresponding pseudo-density $\rho\left(p,q\right)$ has the form
\begin{eqnarray}
W_{\psi}\left(p,q\right) & = & \int d^{N}s\,\psi\left(q+\frac{s}{2}\right)\psi^{*}\left(q-\frac{s}{2}\right)e^{-\frac{i}{\hbar}sp}\,,\label{eq:2-20_def_Wigner_function}
\end{eqnarray}
which is nothing but the Wigner function related to the wave function
$\psi\left(x\right)=\langle x|\psi\rangle$.

Introducing the \emph{Moyal kernel} $K_{{\rm M}}$, we write the formal
solution of the evolution equation (\ref{eq:2-19-Moyal-eq}) in the
form
\begin{eqnarray}
\rho\left(\phi,T\right) & = & \int d^{2N}\phi^\prime
\,K_{{\rm M}}\left(\phi,T;\phi^{\prime},T_{0}\right)\rho\left(\phi^{\prime},T_{0}\right)\,.\label{eq:def-Moyal-kernel}
\end{eqnarray}

By applying the symbol map to equation (\ref{eq:2-12}), the Moyal
kernel $K_{{\rm M}}$ can be expressed in terms of the Feynman kernel.
One obtains
\begin{eqnarray}
K_{{\rm M}}\left(p,q,T;p^{\prime},q^{\prime},T_{0}\right) & = & \left(2\pi\hbar\right)^{-N}\int d^{N}s\int d^{N}s^{\prime}\,e^{\frac{i}{\hbar}\left(s^{\prime}p^{\prime}-sp\right)}{\cal K}\left(s,q,T;s^{\prime},q^{\prime},T_{0}\right)\,,\label{eq:2-22}
\end{eqnarray}
with ${\cal K}$ given by the ``$KK^{*}$-product'':
\begin{eqnarray}
{\cal K}\left(s,q,T;s^{\prime},q^{\prime},T_{0}\right) & \equiv & K\left(q+\frac{s}{2},T;q^{\prime}+\frac{s^{\prime}}{2},T_{0}\right)K\left(q-\frac{s}{2},T;q^{\prime}-\frac{s^{\prime}}{2},T_{0}\right)^{*}\,.\label{eq:2-23}
\end{eqnarray}

Marinov's path integral, on the other hand, while also being a representation of
the Moyal kernel, does not exhibit the pure-state kernel $K$
in any obvious way.

\subsection{Evolving density matrices {\`a} la Marinov}

In his work \cite{MARINOV1991} Marinov starts out directly from Moyal's
evolution equation (\ref{eq:2-19-Moyal-eq}) and solves it in terms
of a functional integral without making any reference to the time
evolution of the \emph{pure} states. Instead, he applies the same
discretization and iteration techniques to eq.~(\ref{eq:2-19-Moyal-eq})
which Feynman had applied to the Schr{\"o}dinger equation in his
derivation of the path integral for the evolution of the wave functions. 

In this manner Marinov arrives at the integral
\begin{eqnarray}
K_{{\rm M}}\left(\phi^{\prime},T;\phi^{\prime\prime},T_{0}\right) & = & \int_{\phi\left(T_{0}\right)=\phi^{\prime\prime}}^{\phi\left(T\right)=\phi^{\prime}}{\cal D}\phi^{a}\left(\cdot\right){\cal D}\xi^{a}\left(\cdot\right)\nonumber \\
 &  & \times \exp\left(-2i\int_{T_{0}}^{T}dt\left\{ \dot{\phi}^{a}\left(t\right)\omega_{ab}\xi^{b}\left(t\right)-\widetilde{H}\left(\phi\left(t\right),\xi\left(t\right)\right)\right\} \right)\label{eq:2-30_def_PI_for_Moyal_kernel}
\end{eqnarray}
which involves the ``bilocal'' Hamiltonian
\begin{eqnarray}
\widetilde{H}\left(\phi,\xi\right) & \equiv & \frac{1}{2\hbar}\left[H\left(\phi-\hbar\xi\right)-H\left(\phi+\hbar\xi\right)\right]\,.\label{eq:2-31_tilde_H-via-phi_and_xi}
\end{eqnarray}
Geometrically speaking, the integration is over parametrized curves
on phase space, $t\mapsto\phi^{a}\left(t\right)$, and vector fields,
$t\mapsto\xi^{a}\left(t\right)$, that are defined along those curves.
While $\phi^{a}$ is subject to boundary conditions, $\xi^{a}$ is
integrated over at the terminal points. 

The continuum notation employed in eq.~(\ref{eq:2-30_def_PI_for_Moyal_kernel})
is only formal clearly. For the exact definition of the path integral
as the ``continuum limit'' of a certain discrete precursor we refer
to the Appendix.

\subsection{Feynman times Feynman implies Marinov}

An alternative derivation of Marinov's path integral, which provides
interesting conceptual insights, has been advocated in ref.~\cite{Gozzi:2020wef}. 

It begins by inserting two copies of the Feynman path integral (\ref{eq:2-13-phase-space-PI})
for $K$ into the eqs.~(\ref{eq:2-22}), (\ref{eq:2-23}) for the
Moyal kernel $K_{{\rm M}}$. Denoting the integration variables by
$\left(p_{+}\left(t\right),x_{+}\left(t\right)\right)\equiv\phi_{+}^{a}\left(t\right)$
in the $K$-, or ``forward''-, or ``ket''-sector and by $\left(p_{-}\left(t\right),x_{-}\left(t\right)\right)\equiv\phi_{-}^{a}\left(t\right)$
in the $K^{*}$-, or ``backward''-, or ``bra''-sector, the (partial)
Fourier transform ${\cal K}$ of $K_{{\rm M}}$ reads, formally,
\begin{eqnarray}
{\cal K}\left(s,q,T;s^{\prime},q^{\prime},T_{0}\right) & = & \int_{x_{+}\left(T_{0}\right)=q^{\prime}+\frac{s^{\prime}}{2}}^{x_{+}\left(T\right)=q+\frac{s}{2}}{\cal D}x_{+}\left(\cdot\right)\int{\cal D}p_{+}\left(\cdot\right)\int_{x_{-}\left(T_{0}\right)=q^{\prime}-\frac{s^{\prime}}{2}}^{x_{-}\left(T\right)=q-\frac{s}{2}}{\cal D}x_{-}\left(\cdot\right)\int{\cal D}p_{-}\left(\cdot\right)\nonumber \\
 &  & \times \exp\left(\frac{i}{\hbar}\int_{T_{0}}^{T}dt\left\{ p_{+}\dot{x}_{+}-p_{-}\dot{x}_{-}-H\left(p_{+},x_{+}\right)+H\left(p_{-},x_{-}\right)\right\} \right)\,.\label{eq:2-32-Y_kernel-in-phase-space}
\end{eqnarray}
Eq.~(\ref{eq:2-32-Y_kernel-in-phase-space}) amounts to the integration
over paths on the doubled phase space ${\cal M}\times{\cal M}$. 

Next we replace the above integration variables $\phi_{+}^{a}\left(t\right),\phi_{-}^{a}\left(t\right)$
by new ones, $\phi^{a}\left(t\right)$ and $\xi^{a}\left(t\right)$.
They are defined as
\begin{eqnarray}
\phi^{a} & = & \frac{1}{2}\left(\phi_{+}^{a}+\phi_{-}^{a}\right)\,,\nonumber \\
\xi^{a} & = & \frac{1}{2\hbar}\left(\phi_{+}^{a}-\phi_{-}^{a}\right)\,,\label{eq:2-33-phi_symm_and_xi_antisymm}
\end{eqnarray}
and the inverse transformation reads
\begin{eqnarray}
\phi_{\pm}^{a} & = & \phi^{a}\pm\hbar\xi^{a}\,.\label{eq:2-200-phi_pm_via_phi_and_xi}
\end{eqnarray}

Now the main effort in proving Marinov's path integral formula consists
in showing that when the change of variables (\ref{eq:2-33-phi_symm_and_xi_antisymm})
is properly implemented, eq.~(\ref{eq:2-32-Y_kernel-in-phase-space})
when inserted into (\ref{eq:2-22}) is equivalent to the path integral
for $K_{{\rm M}}$ as given in eq.~(\ref{eq:2-30_def_PI_for_Moyal_kernel}). 

While this is deceptively easy to do if one uses formal continuum
notations, and actually has been discussed at this level in ref.~\cite{Gozzi:2020wef},
it goes without saying that in order to give a meaningful proof, all
functional integrals involved must be dealt with in terms of their
discrete regularizations which actually define them. Starting out
from the two copies of the regularized path integral for $K$, we
must derive the discrete counterpart of (\ref{eq:2-32-Y_kernel-in-phase-space})
and then try to find a well defined, finite dimensional change of
integration variables for non-zero lattice spacing, such that the
result is equivalent to the sequence of the $n$-fold integrals that
defines Marinov's path integral.

In the Appendix we show that such a transformation can indeed be found
and leads to a new derivation of Marinov's formula. The non-trivial
point hereby is that the discrete analog of eq.~(\ref{eq:2-33-phi_symm_and_xi_antisymm})
must be tailored in a very specific way. Indeed, one may think that
the Fourier transforms present in (\ref{eq:2-22}) hamper expressing
the Moyal kernel $K_{{\rm M}}$ as a discretized path integral over
$\phi^{a}$ and $\xi^{a}$ with the formers having fixed endpoints.
However, by adopting the pre-point prescription for the Feynman kernels,
it is possible to perform a change of variables at the discretized
level that, together with a couple of integrations, allows one to
write the path integral in the desired form.

\section{The random surface representation} \label{sec:rnd_surface_representation}

In this section we are going to show that the Marinov integral (\ref{eq:2-30_def_PI_for_Moyal_kernel})
admits a natural interpretation as a ``sum'' over 2D random surfaces
in phase space. 

\subsection{Ruled surfaces}

\begin{figure}
\captionsetup{justification=raggedright}
\includegraphics[scale=0.35]{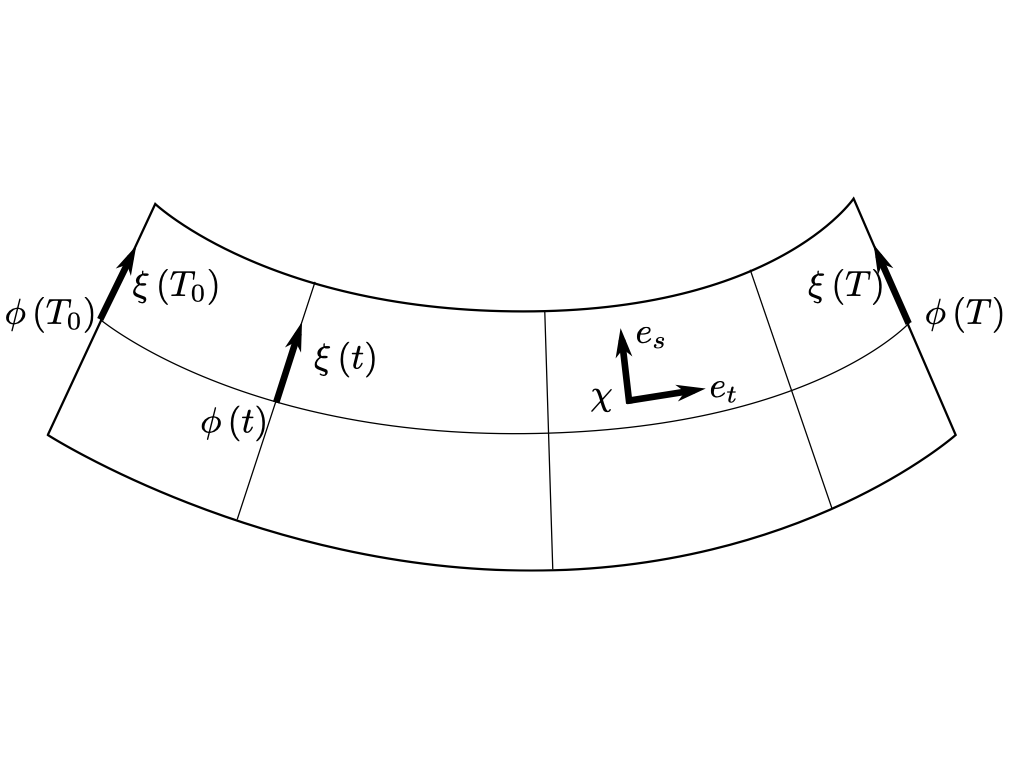}

\caption{The ruled surface in phase space defined by eq.~(\ref{eq:3-35}).
Shown is the directrix $\phi\left(t\right)$, the vector field $\xi\left(t\right)$
along this curve, and some of the surface's straight-line generators
tangent to $\xi\left(t\right)$.\label{fig:1}}

\end{figure}

Given an arbitrary trajectory on ${\cal M}$, $t\mapsto\phi^{a}\left(t\right)$,
and a vector field $t\mapsto\xi^{a}\left(t\right)$ defined along
this trajectory, let us consider the map
\begin{eqnarray}
  \chi:\left[T_{0},T\right]\times\left[-1,1\right] & \rightarrow & {\cal M}\nonumber \\
  \left(t,s\right) & \mapsto & \chi^{a}\left(t,s\right)\equiv\phi^{a}\left(t\right)+s\hbar \, \xi^{a}\left(t\right)\,.\label{eq:3-35}
\end{eqnarray}
We interpret (\ref{eq:3-35}) as the parametrization of a certain
surface $\Sigma$ in ${\cal M}$. It is not an entirely generic one,
but rather a \emph{ruled surface} \cite{RS1,RS2}. The curve $\phi^{a}\left(t\right)$
plays the role of its \emph{directrix}, while $\xi^{a}\left(t\right)$
supplies the information about the direction of the \emph{generating
lines} $s\mapsto\chi\left(t,s\right)$, $t$ fixed. In fact, by its
very definition, through every point of a ruled surface there is at
least one straight line lying on the surface, see Figure \ref{fig:1}
for an illustration.

In the following the new parameter $s\in\left[-1,1\right]$ will enjoy
the same logical status as $t$. In particular, the corresponding
partial derivatives of $\chi$ provide us with two linearly independent
vectors tangent to $\Sigma$, $e_{s}$ and $e_{t}$, respectively\footnote{We abbreviate $\partial_{s}\equiv\partial/\partial s$, $\partial_{t}\equiv\partial/\partial t$,
and sometimes also use a dot to indicate $t$-derivatives.}
\begin{eqnarray}
e_{s}: & \,  & \partial_{s}\chi^{a}\left(t,s\right)=\hbar \, \xi^{a}\left(t\right)\label{eq:3-40}\\
e_{t}: & \, & \partial_{t}\chi^{a}\left(t,s\right)=\dot{\phi}^{a}\left(t\right)+\hbar s \, \dot{\xi}^{a}\left(t\right)\,.\nonumber 
\end{eqnarray}
See Figure \ref{fig:1} for a schematic sketch.

\begin{figure}
\captionsetup{justification=raggedright}
\includegraphics[scale=0.35]{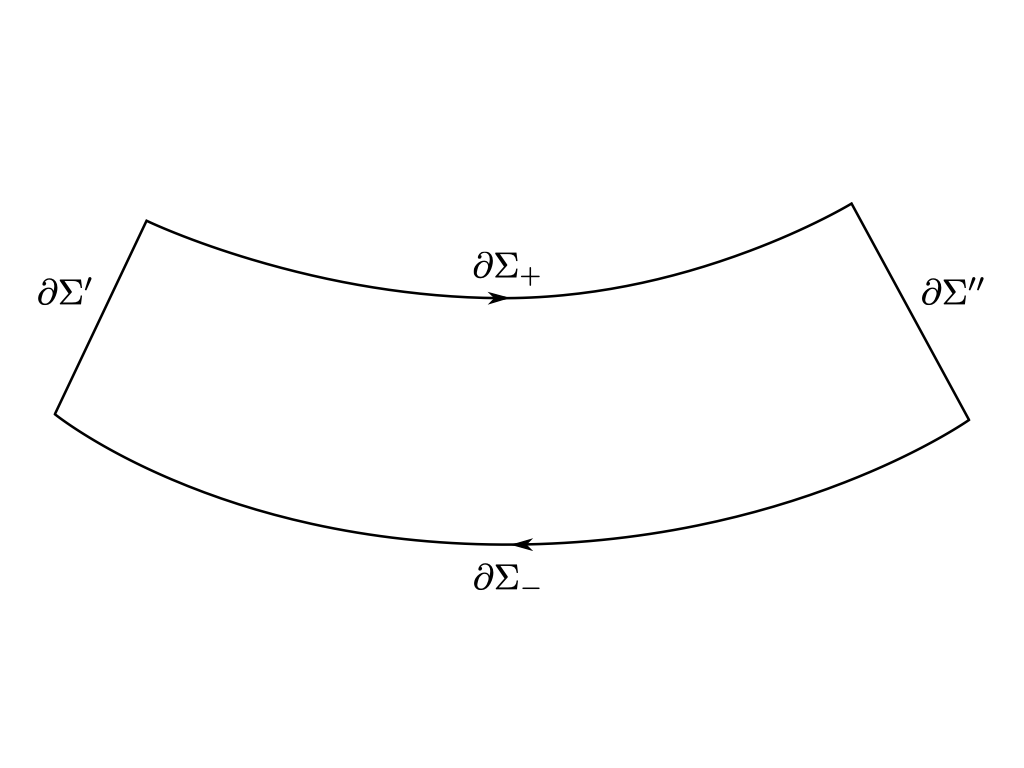}

\caption{The four parts of the boundary $\partial\Sigma$ displayed by the
surface sketched in Figure \ref{fig:1}.\label{fig:2}}

\end{figure}

The surface $\Sigma$ has a non-empty boundary, $\partial\Sigma$.
It consists of 4 components, a parametrization of which follows from
$\chi\left(t,s\right)$ by fixing one of the two parameters, according
to
\begin{eqnarray}
\partial\Sigma^{\prime} & : & t=T_{0}\nonumber \\
\partial\Sigma^{\prime\prime} & : & t=T\nonumber \\
\partial\Sigma_{+} & : & s=+1\label{eq:3-41}\\
\partial\Sigma_{-} & : & s=-1\,.\nonumber 
\end{eqnarray}
Thus, $\partial\Sigma=\partial\Sigma^{\prime}\cup\partial\Sigma^{\prime\prime}\cup\partial\Sigma_{+}\cup\partial\Sigma_{-}$,
see Figure \ref{fig:2} for an illustration.

\begin{figure}
\captionsetup{justification=raggedright}
\includegraphics[scale=0.35]{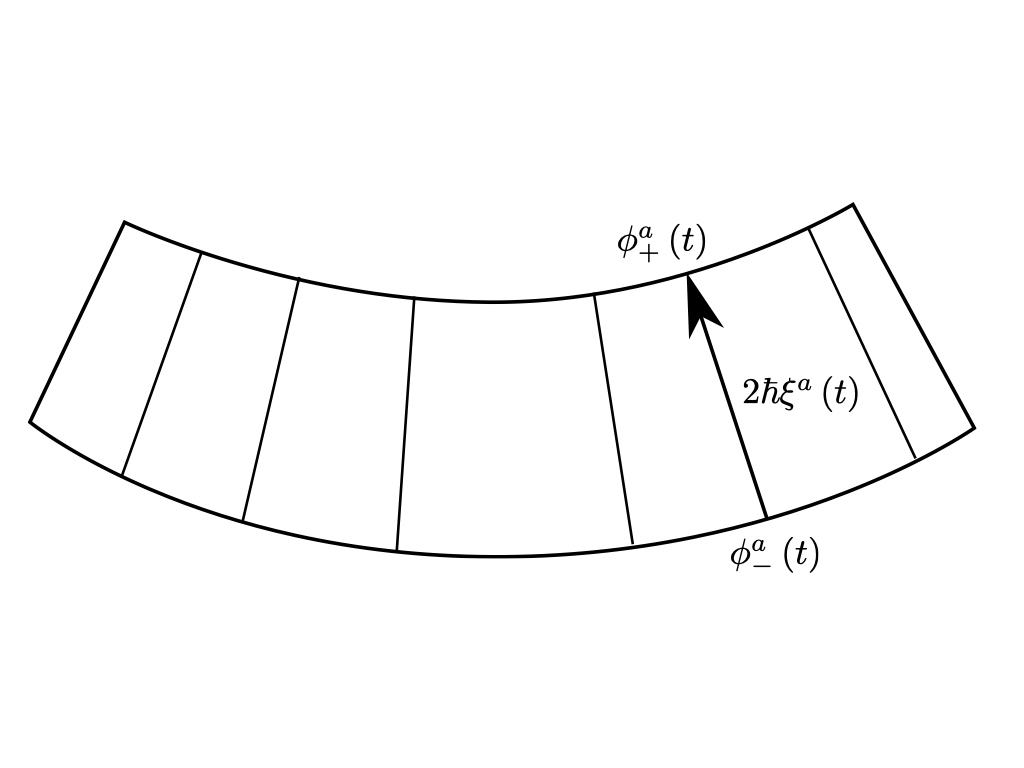}

\caption{The ruled surface defined by the trajectories $\phi_{+}^{a}\left(\cdot\right)$
and $\phi_{-}^{a}\left(\cdot\right)$, which are related to the ``ket''-
and the ``bra''-sector, respectively. At time $t$, they are connected by the vector
$2\hbar\xi^{a}\left(t\right)$.\label{fig:3}}

\end{figure}

Let us use for a moment the ``bra'' and ``ket'' variables
$\left(\phi_{-}^{a},\phi_{+}^{a}\right)$ instead of $\phi^{a}$ and
$\xi^{a}$. If we insert (\ref{eq:2-33-phi_symm_and_xi_antisymm})
into (\ref{eq:3-35}) the parametrization of $\Sigma$ assumes the
form
\begin{eqnarray}
\chi^{a}\left(t,s\right) & = & \frac{1}{2}\left(1-s\right)\phi_{-}^{a}\left(t\right)+\frac{1}{2}\left(1+s\right)\phi_{+}^{a}\left(t\right)\,.\label{eq:3-410}
\end{eqnarray}
We see that when $s$-increases from $-1$ to $+1$, the point $\chi^{a}\left(t,s\right)$,
$t$ fixed, moves on a straight line of finite extension from $\phi_{-}^{a}\left(t\right)$
to $\phi_{+}^{a}\left(t\right)$. It is a generator of $\Sigma$ and
has the direction of
\begin{eqnarray}
2\hbar\,\xi^{a}\left(t\right) & \equiv & \phi_{+}^{a}\left(t\right)-\phi_{-}^{a}\left(t\right)\,.\label{eq:3-411}
\end{eqnarray}
If we now increase $t$ from $T_{0}$ to $T$, we can visualize the
resulting surface as the ``worldsheet'' swept out by a ``stiff
string'' in phase space. It is guided by the two directices $\phi_{+}^{a}\left(t\right)$
and $\phi_{-}^{a}\left(t\right)$, which define the boundaries $\partial\Sigma_{+}$
and $\partial\Sigma_{-}$, respectively, see Figure \ref{fig:3}.

\subsection{From paths to surfaces}

Next we are going to reformulate Marinov's path integral
\begin{eqnarray}
K_{{\rm M}}\left(\phi^{\prime},T;\phi^{\prime\prime},T_{0}\right) & = & \int_{\phi\left(T_{0}\right)=\phi^{\prime\prime}}^{\phi\left(T\right)=\phi^{\prime}}{\cal D}\phi^{a}\left(\cdot\right){\cal D}\xi^{a}\left(\cdot\right)\exp\Bigr(-2iS_{{\rm M}}\left[\phi\left(\cdot\right),\xi\left(\cdot\right)\right]\Bigr)\,,\label{eq:3-Mari}
\end{eqnarray}
in terms of the surfaces parametrized by $\chi^{a}\left(t,s\right)$.
In eq.~(\ref{eq:3-Mari}) we introduced the Marinov action functional
\begin{eqnarray}
S_{{\rm M}}\left[\phi\left(\cdot\right),\xi\left(\cdot\right)\right] & \equiv & \int_{T_{0}}^{T}dt\left\{ \dot{\phi}^{a}\left(t\right)\omega_{ab}\xi^{b}\left(t\right)-\widetilde{H}\left(\phi\left(t\right),\xi\left(t\right)\right)\right\} \,,\label{eq:3-46}
\end{eqnarray}
which, as we discussed above, arises according to
\begin{eqnarray}
S_{{\rm M}}\left[\phi\left(\cdot\right),\xi\left(\cdot\right)\right] & = & -\frac{1}{2\hbar}\Bigr\{ S\left[\phi_{+}\right]-S\left[\phi_{-}\right]\Bigr\}\,,\label{eq:3-SMS}
\end{eqnarray}
where $S=\int dt\,\left\{ p\dot{x}-H\right\} $ denotes the ordinary first
order action.

\noindent {\bf{(1) Helpful identities.}} 
Below we shall need the following two relations:
\begin{eqnarray}
\frac{1}{2\hbar}\int_{-1}^{+1}ds\,\omega_{ab} \, \partial_{t}\chi^{a}\left(t,s\right)\partial_{s}\chi^{b}\left(t,s\right) & = & \dot{\phi}^{a}\omega_{ab}\xi^{b}\,,\label{eq:3-42}\\
\frac{1}{2\hbar}\int_{-1}^{+1}ds\,\partial_{s}\chi^{a}\left(t,s\right)\partial_{a}H\left(t,s\right) & = & -\widetilde{H}\left(\phi\left(t\right),\xi\left(t\right)\right)\,.\label{eq:3-43}
\end{eqnarray}
Eq.~(\ref{eq:3-42}) follows straightforwardly by noting that
\begin{eqnarray}
 &  & \int_{-1}^{+1}ds\,\omega_{ab} \, \partial_{t}\chi^{a}\left(t,s\right)\partial_{s}\chi^{b}\left(t,s\right)\,=\,\int_{-1}^{+1}ds\,\omega_{ab}\left[\dot{\phi}^{a}+s\hbar\dot{\xi}^{a}\right]\hbar\xi^{b}\,=\,\nonumber \\
 &  & \int_{-1}^{+1}ds\,\omega_{ab} \, \dot{\phi}^{a}\hbar\xi^{b}+\hbar^{2}\int_{-1}^{+1}ds\,s\,\dot{\xi}^{a}\omega_{ab}\xi^{b}\,=\,\left(2\hbar\right)\dot{\phi}^{a}\omega_{ab}\xi^{b}\,.\label{eq:3-44}
\end{eqnarray}
Here the $O\left(\hbar^{2}\right)$ term vanishes as a consequence
of the $s$-integration, $\int_{-1}^{+1}ds\,s=0$, whereas the $O\left(\hbar\right)$
term is non-zero: $\int_{-1}^{+1}ds=2$.

The relation (\ref{eq:3-43}) is proved by the following steps:
\begin{eqnarray}
 &  & \int_{-1}^{+1}ds\,\partial_{s}\chi^{a}\partial_{a}H\left(\chi\right)\,=\,\nonumber \\
 &  & \,=\,\int_{-1}^{+1}ds\,\partial_{s}\left[\phi^{a}+s\hbar\xi^{a}\right]\partial_{a}H\left(\phi+s\hbar\xi\right)\nonumber \\
 &  & \,=\,\int_{-1}^{+1}ds\,\frac{d}{ds}H\left(\phi+s\hbar\xi\right)\nonumber \\
 &  & \,=\,H\left(\phi+\hbar\xi\right)-H\left(\phi-\hbar\xi\right)\nonumber \\
 &  & \,=\, -\left(2\hbar\right)\widetilde{H}\left(\phi , \xi\right)\,.\label{eq:3-45}
\end{eqnarray}

\noindent {\bf{(2) The action $S_{{\rm M}}\left[\chi\right]$ for surfaces.}} 
By virtue of the identities (\ref{eq:3-42})
and (\ref{eq:3-43}) the action functional that appears under Marinov's
path integral, i.e., $S_{{\rm M}}\left[\chi\right]\equiv S_{{\rm M}}\left[\phi,\xi\right]$
given in eq.~(\ref{eq:3-46}), is seen to admit the following rather
suggestive representation entirely in terms of the parametrization
$\chi$:
\begin{equation}
\boxed{
S_{{\rm M}}\left[\chi\right] = \frac{1}{2\hbar}\int_{-1}^{+1}ds\int_{T_{0}}^{T}dt\Bigr\{\omega_{ab}\partial_{t}\chi^{a}\left(t,s\right)\partial_{s}\chi^{b}\left(t,s\right)+\partial_{s}\chi^{a}\left(t,s\right)\partial_{a}H\left(\chi\left(t,s\right)\right)\Bigr\}\,.
}\label{eq:3-47}
\end{equation}

The action $S_{{\rm M}}$ as displayed in eq.~(\ref{eq:3-47}) has a clear geometrical
interpretation. We denote the first and second term on the RHS of
(\ref{eq:3-47}) by $S_{{\rm M}}^{\left(0\right)}$ and $S_{{\rm M}}^{\left(H\right)}$,
respectively, and discuss them in turn.

\noindent {\bf{(3) The area term $S_{{\rm M}}^{\left(0\right)}$.}} 
Let $\omega=\frac{1}{2}\omega_{ab}d\phi^{a}\wedge d\phi^{b}$
denote the symplectic $2$-form on ${\cal M}$, and let $\Sigma$
be an arbitrary (not necessarily ruled, for this matter) 2D surface
in phase space, parametrized by $\chi:{\cal U}\rightarrow{\cal M}$,
$\left(t,s\right)\mapsto\chi^{a}\left(t,s\right)$ for some ${\cal U}\subset\mathbb{R}^{2}$.
Then the $2$-form $\omega$ evaluated on the pertinent tangent vectors
$e_{s}=e_{s}^{a}\partial_{a}=\partial_{s}\chi^{a}\partial_{a}$ and
$e_{t}=e_{t}^{a}\partial_{a}=\partial_{t}\chi^{a}\partial_{a}$, respectively,
reads $\omega\left(e_{t},e_{s}\right)=\omega_{ab}\partial_{t}\chi^{a}\partial_{s}\chi^{b}$.
Hence the intrinsically defined integral $\int_{\Sigma}\omega$ of
the $2$-form $\omega$ over the surface $\Sigma$ is given by
\begin{eqnarray}
\int_{\Sigma}\omega & = & \int_{{\cal U}}ds\,dt\,\omega\left(e_{t},e_{s}\right)\,=\,\int_{{\cal U}}ds\,dt\,\omega_{ab} \, \partial_{t}\chi^{a}\partial_{s}\chi^{b}\,.\label{eq:3-50}
\end{eqnarray}

Obviously the first term of the action functional $S_{{\rm M}}\equiv S_{{\rm M}}^{\left(0\right)}+S_{{\rm M}}^{\left(H\right)}$
in (\ref{eq:3-47}) has exactly this structure, with ${\cal U}=\left[T_{0},T\right]\times\left[-1,1\right]$,
and so we may conclude that
\begin{eqnarray}
S_{{\rm M}}^{\left(0\right)} & = & \frac{1}{2\hbar}\int_{\Sigma}\omega\,.\label{eq:3-51}
\end{eqnarray}
Thus $S_{{\rm M}}^{\left(0\right)}$ is nothing but the symplectic
area of $\Sigma$ in units of $2\hbar$. It is invariant both under
symplectic diffeomorphism of ${\cal M}$ (canonical transformations)
and under reparametrizations of $\Sigma$.

We also note in passing that the parametrized form of $S_{{\rm M}}^{\left(0\right)}$
acquires a more symmetric appearance introducing unified ``world
sheet coordinates'' on $\Sigma$, namely $\left(\sigma^{\alpha}\right)\equiv\left(\sigma^{1},\sigma^{2}\right)=\left(t,s\right)$:
\begin{eqnarray}
S_{{\rm M}}^{\left(0\right)} & = & \frac{1}{4\hbar}\int_{{\cal U}}d^{2}\sigma\,\varepsilon^{\alpha\beta}\omega_{ab}\partial_{\alpha}\chi^{a}\left(\sigma\right)\partial_{\beta}\chi^{b}\left(\sigma\right)\,.\label{eq:3-55-strings}
\end{eqnarray}
Here $\alpha,\beta\in\left\{ 1,2\right\} $, $\partial_{\alpha}\equiv\partial/\partial\sigma^{\alpha}$,
and the 2D epsilon tensor $\varepsilon^{\alpha\beta}$ is defined
such that $\varepsilon^{12}=1$.

\noindent {\bf{(4) The Hamiltonian term $S_{{\rm M}}^{\left(H\right)}$.}} 
Contrary to $S_{{\rm M}}^{\left(0\right)}$,
which is of a universal and purely geometrical nature, the second
term of the Marinov action, $S_{{\rm M}}^{\left(H\right)}$, depends
on the Hamiltonian and reflects the specific dynamics of the physical
system under consideration. Geometrically speaking, it involves the
differential of the (ordinary!) Hamiltonian, $dH=\partial_{a}H\left(\phi\right)d\phi^{a}$,
evaluated on $e_{s}$:
\begin{eqnarray}
S_{{\rm M}}^{\left(H\right)} & = & \frac{1}{2\hbar}\int_{{\cal U}}ds\,dt\,\partial_{s}\chi^{a}\left(t,s\right)\partial_{a}H\left(\chi\left(t,s\right)\right)\nonumber \\
 & = & \frac{1}{2\hbar}\int_{{\cal U}}ds\,dt\,dH\left(e_{s}\right)\,.\label{eq:3-60}
\end{eqnarray}
This part of the action, too, is invariant under symplectic diffeomorphisms
of ${\cal M}$. It is not invariant though under reparametrizations
of the surface $\Sigma$.

It is natural to interpret $dH\left(\phi\right)$ as an ``external
field'' to which ``free'' surfaces governed by $S_{{\rm M}}^{\left(0\right)}$
couple via the interaction term $S_{{\rm M}}^{\left(H\right)}$.

We saw that for the free part $S_{{\rm M}}^{\left(0\right)}$ the
geometrically most natural interpretation is that of a surface integral.
For $S_{{\rm M}}^{\left(H\right)}$ the situation is different. Due
to its lack of reparametrization invariance the representation (\ref{eq:3-60})
in terms of a double integration, while correct, appears somewhat
artificial. In fact, the most natural way of thinking about $S_{{\rm M}}^{\left(H\right)}$
is in terms of an external field which couples to $\Sigma$ \emph{at
its boundary}. Rather that using the identity (\ref{eq:3-43}), we
note that $\chi^{a}\left(\pm1,t\right)=\phi^{a}\pm\hbar\xi^{a}$,
and employ (\ref{eq:2-31_tilde_H-via-phi_and_xi}) to obtain
\begin{eqnarray}
-\left(2\hbar\right)\int_{T_{0}}^{T}dt\,\widetilde{H}\left(\phi\left(t\right),\xi\left(t\right)\right) & = & \int_{T_{0}}^{T}dt\,\left\{ H\left(\chi\left(1,t\right)\right)-H\left(\chi\left(-1,t\right)\right)\right\} \nonumber \\
 & = & \int_{\partial\Sigma_{+}}H\,dt+\int_{\partial\Sigma_{-}}H\,dt\,.\label{eq:3-80}
\end{eqnarray}
Note the signs of the integrals in the last line of eq.~(\ref{eq:3-80}).
They are fixed by the orientation we adopted for the boundary components
$\partial\Sigma_{+}$ and $\partial\Sigma_{-}$, respectively, see
Figure \ref{fig:2}. 
By eq.~(\ref{eq:3-80}) the ``external field'' $H\left(\phi\right)$,
a scalar on the ambient space, is seen to couple to the ruled surfaces
$\Sigma$ along their two ``spatial'' boundaries:
\begin{eqnarray}
S_{{\rm M}}^{\left(H\right)} & = & \frac{1}{2\hbar}\int_{\partial\Sigma_{+}\cup\partial\Sigma_{-}}Hdt\,.\label{eq:3-81}
\end{eqnarray}

\noindent {\bf{(5) The Moyal-Marinov kernel as a sum over surfaces.}} 
Returning now to the kernel $K_{{\rm M}}$,
we find that it can be represented it by the following functional
integral over ruled surfaces $\Sigma$:
\begin{eqnarray}
 & \boxed{K_{{\rm M}}\left(\phi^{\prime},T;\phi^{\prime\prime},T_{0}\right)=\int{\cal D}\chi\,\exp\left\{ -\frac{i}{\hbar}\left(\int_{\Sigma}\omega+\int_{\partial\Sigma_{+}\cup\partial\Sigma_{-}}Hdt\right)\right\} \,.}\label{eq:3-85}
\end{eqnarray}
Hereby the integration over the parametrizations $\chi\left(s,t\right)$
is subject to the conditions
\begin{eqnarray*}
 & \boxed{\chi\left(0,T_{0}\right)=\phi^{\prime}\,,\,\chi\left(0,T\right)=\phi^{\prime\prime}\,.}
\end{eqnarray*}
From the surface perspective, only the respective \emph{central points}
of the straight-line boundaries $\partial\Sigma^{\prime}$ and $\partial\Sigma^{\prime\prime}$
get fixed by the boundary conditions. Their \emph{length} and \emph{orientation}
in phase space are arbitrary instead, i.e., integrated over,
the reason being that the original $\xi$-integrations had been unconstrained
at $t=T_0$ and $t=T$.

\subsection{A ``string theory'' on phase space}

In order to gain a certain degree of intuitive understanding of Marinov's
path integral it may be helpful to think of the theory of surfaces
defined by eq.~(\ref{eq:3-85}) as a kind of ``string theory''.
Hence a comparison with relativistic (super-)strings suggests itself
at this point. 

Both frameworks deal with $2$-dimensional ``worldsheets'' $\Sigma$
in a higher dimensional ambient space, interpreted as spacetime in
relativistic string theory, and as phase space in the present case. 

The Nambu-Goto action, protoptype of an action functional for free
relativistic strings, evaluates the \emph{Riemannian} surface area
of the worldsheet according to the metric which the embedding induces
from the spacetime metric. In the case at hand, the free action $S_{{\rm M}}^{\left(0\right)}$
is precisely the \emph{symplectic} analogue of this area, $\int_{\Sigma}\omega$,
which similarly involves a pullback from ${\cal M}$ to $\Sigma$,
namely that of the canonical $2$-form $\omega$. 

Furthermore, when written
in the style of eq.~(\ref{eq:3-55-strings}), 
the {\it free} action of the symplectic case, $S_{{\rm M}}^{\left(0\right)}$,
happens to have the same structure 
as the standard {\it interaction} term that couples a background
$2$-form field $B_{\mu\nu}$ to the worldsheet of a superstring 
\cite{QFT_and_strings_for_math}.
Thereby, the role of $B_{\mu\nu}$ is taken over by $\omega_{ab}$ now. 

Abelian and non-Abelian $1$-form gauge-fields $A_{\mu}$, on the
other hand, naturally couple to open relativistic strings on the boundary
of the worldsheet; in a rather similar fashion the interaction term
$S_{{\rm M}}^{\left(H\right)}$ couples to the surfaces $\Sigma$
to the ``external field'' $\partial_{a}H$ which is responsible
for a non-trivial dynamics.


\section{Summing classical surfaces} \label{sec:sum_classical_surf}

\begin{figure}
\captionsetup{justification=raggedright}
\includegraphics[scale=0.35]{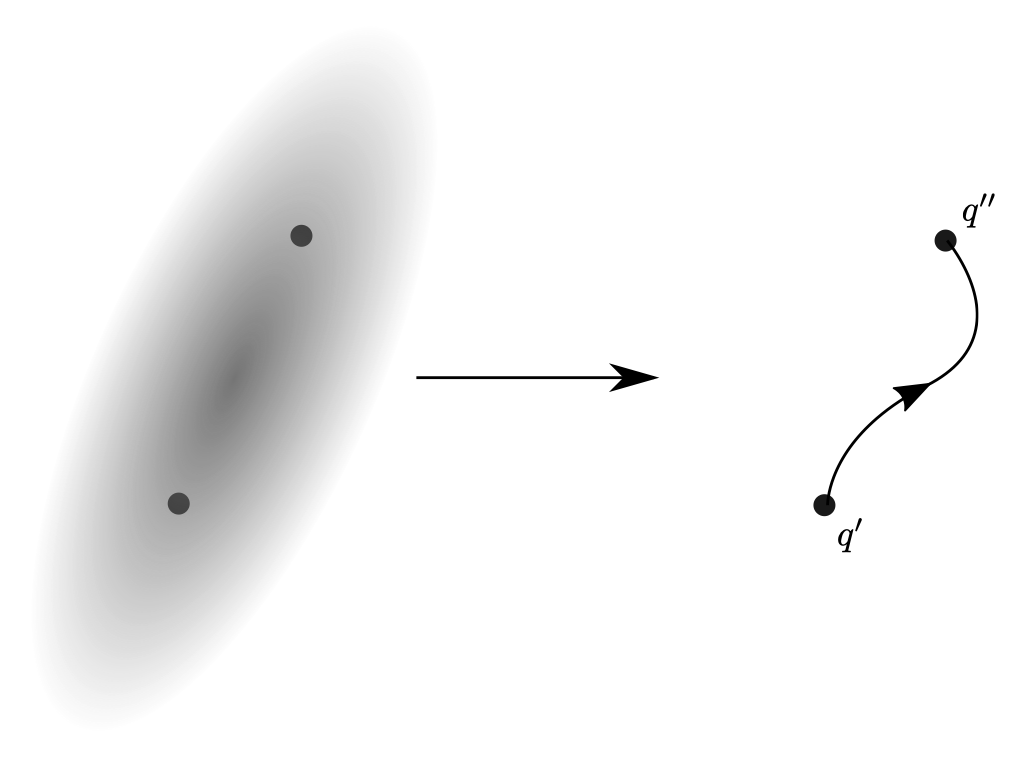}
\caption{In the limit $\hbar\rightarrow0$, a cloud of paths on configuration space
connecting $q^{\prime}$ to $q^{\prime\prime}$ shrinks onto a single classical particle trajectory.\label{fig:6a}}
\end{figure}
The semi-classical limit of Feynman-type path integrals is commonly
interpreted in the heuristic way shown in Figure \ref{fig:6a}:
While in the quantum regime (``$\hbar$ large'') the transition
amplitude between two points $q^{\prime}$ to $q^{\prime\prime}$ receives essential contributions from
a huge cloud of mostly non-classical paths, this cloud shrinks in
the limit $\hbar\rightarrow0$ to an infinitesimally thin tube that
surrounds the classical configuration space trajectory connecting
those points.

\noindent {\bf{(1)}} For the surface formulation of the Marinov path
integral, this behavior in the classical limit translates to the following
heuristics. Deep in the quantum regime, those surfaces $\Sigma$
which make the dominant contribution to $K_{\rm M}$ are comparatively
large phase space objects.
And typically
their projections on configuration space, denoted $\Pi_{{\rm con}}\Sigma$
in the following, are relatively large, too. 

In fact, by virtue of eqs.~(\ref{eq:3-35}) and (\ref{eq:3-411})
the ``size'' of $\Sigma$ can be estimated in terms of $\left\Vert \hbar\xi\right\Vert $,
the Euclidean norm of the response field. 
In the quantum domain, $2\hbar\xi=\phi_{+}-\phi_{-}$
is indeed large typically, the reason being that essential contributions
to the integral can come from non-classical boundary $\phi_{+}\left(t\right)$
and $\phi_{-}\left(t\right)$ which deviate strongly from one another.

\noindent {\bf{(2)}} 
If, for $\hbar\rightarrow0$, both $\phi_{+}\left(t\right)$ and $\phi_{-}\left(t\right)$
approach \emph{the same} classical phase space trajectory, 
the typical size $\left\Vert \phi_{+}-\phi_{-}\right\Vert $ of the relevant surfaces
(and of their configuration space projections) is bound to shrink. 
As a result, the classical
limit is dominated by degenerate, basically $1$-dimensional objects, namely
ordinary trajectories rather than ``worldsheets''.

\begin{figure}
\captionsetup{justification=raggedright}
\includegraphics[scale=0.35]{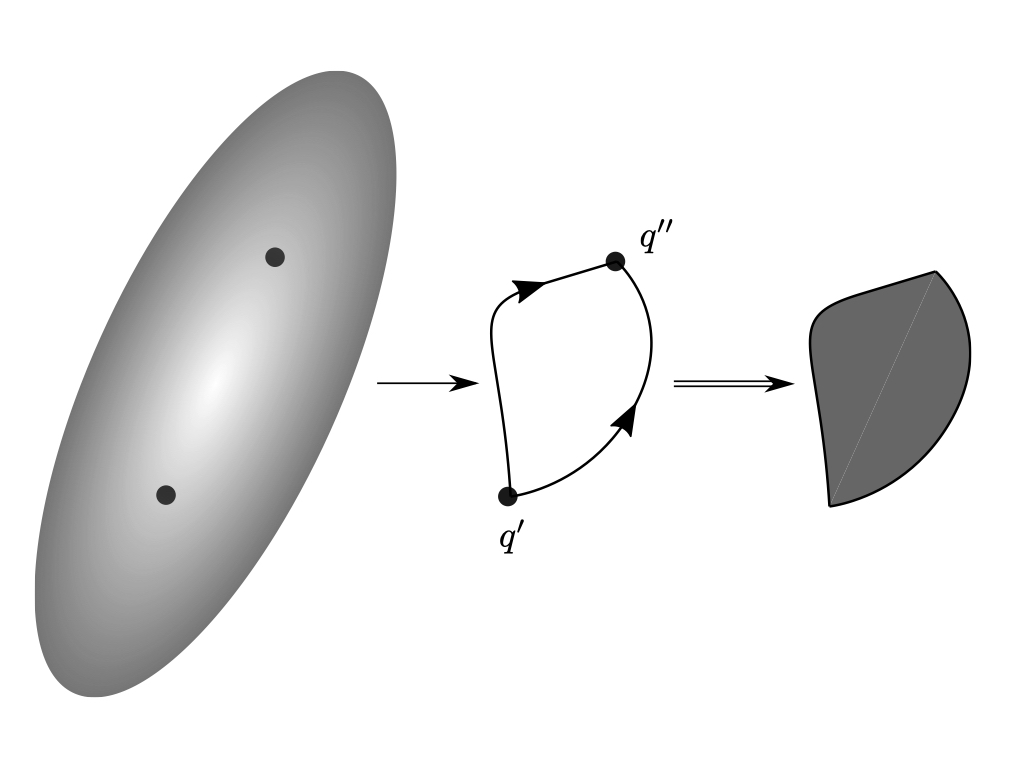}
\caption{For $\hbar\rightarrow0$, the cloud of quantum paths collapses onto
two different trajectories from $q^{\prime}$ to $q^{\prime\prime}$.
In this case the sum-over-surfaces receives essential contributions
from very large surfaces $\Sigma$ also, possibly with macroscopic
projections $\Pi_{{\rm con}}\Sigma$ even. \label{fig:6b}}

\end{figure}

\noindent {\bf{(3)}} However, this picture changes in an essential
way should there exist more that one classical phase space trajectory 
whose projection on configuration space connects the two prescribed
terminal points $q^{\prime}$ and $q^{\prime\prime}$, respectively.
Then, as illustrated in Figure \ref{fig:6b},
even very large surfaces can survive the limit $\hbar\rightarrow0$
if their directrices $\phi_{+}\left(t\right)$ and $\phi_{-}\left(t\right)$
approach \emph{different} classical trajectories.
 The magnitude of $\hbar\xi\left(t\right)$ which determines the length
of the corresponding ``string'' is dictated by the vectorial difference
of the two classical paths then; under appropriate conditions it may
assume macroscopic values even. 

If surfaces of this kind make an essential contribution to the transition
amplitude, this leads to interference phenomena which have a spatially
extended, and wave-like appearence often. Hereby the correlations
between distant points are mediated by worldsheets that are spread
out in between two classical particle trajectories.

In the following we look more closely at this possibility and present
some illustrative examples. 

\subsection{Semiclassical sum over surfaces}

To keep the kinematics simple, let us restrict the arguments of the
Marinov integral in such a way that it evaluates the probability density
\begin{eqnarray}
\left|K\left(q^{\prime\prime},T;q^{\prime},T_{0}\right)\right|^{2} & = & {\cal K}\left(0,q^{\prime\prime},T;0,q^{\prime},T_{0}\right)\,.\label{eq:4-100}
\end{eqnarray}

We assume furthermore that in the kinematic regime of interest the
Feynman kernel
\begin{eqnarray}
K\left(q^{\prime\prime},T;q^{\prime},T_{0}\right) & = & \int_{\varphi^{q}\left(T_{0}\right)=q^{\prime}}^{\varphi^{q}\left(T\right)=q^{\prime\prime}}{\cal D}\varphi^{a}\left(\cdot\right)\,e^{\frac{i}{\hbar}S\left[\varphi^{a}\left(\cdot\right)\right]}\label{eq:4-101}
\end{eqnarray}
is well approximated by the lowest order of the semiclassical expansion:
\begin{eqnarray}
K\left(q^{\prime\prime},T;q^{\prime},T_{0}\right) & \approx & \sum_{\alpha={\rm I},{\rm II},{\rm III},\cdots}\exp\left(\frac{i}{\hbar}S\left[\varphi_{{\rm SP}}^{\left(\alpha\right)}\left(\cdot\right)\right]\right)\,.\label{eq:4-102}
\end{eqnarray}
In writing down eq.~(\ref{eq:4-101}) we combined the integration
variables $p\left(t\right),x\left(t\right)$ from (\ref{eq:2-13-phase-space-PI})
in the phase space path $\varphi^{a}\left(t\right)\equiv\left(\varphi^{p}\left(t\right),\varphi^{q}\left(t\right)\right)\equiv\left(p\left(t\right),q\left(t\right)\right)$,
and we denoted the first order form of the action by $S\left[\varphi\right]$. 

In the approximation (\ref{eq:4-102}), the sum runs over the set
of stationary points $\left\{ \varphi_{{\rm SP}}^{\left(\alpha\right)}\right\} _{\alpha={\rm I},\,{\rm II},\,{\rm III},\cdots}$.
They satisfy Hamilton's equation
\begin{eqnarray}
\frac{\delta S}{\delta\varphi^{a}\left(t\right)}\left[\varphi_{{\rm SP}}^{\left(\alpha\right)}\right] & = & 0\label{eq:4-103}
\end{eqnarray}
subject to the boundary conditions
\begin{eqnarray}
\varphi^{q}\left(T_{0}\right)=q^{\prime} & ,\, & \varphi^{q}\left(T\right)=q^{\prime\prime}\,,\label{eq:4-104}
\end{eqnarray}
the momenta $\varphi^{p}$ being arbitrary at the terminal points.

Inserting (\ref{eq:4-102}) into (\ref{eq:4-100}) for the factors
$K$- and $K^{*}$, respectively, we obtain the following approximation
for ${\cal K}$ at vanishing $s$-arguments:
\begin{eqnarray}
{\cal K}\left(0,q^{\prime\prime},T;0,q^{\prime},T_{0}\right) & = & \sum_{\alpha,\beta={\rm I},{\rm II},{\rm III}\cdots}e^{\frac{i}{\hbar}S\left[\phi_{+}=\varphi_{{\rm SP}}^{\left(\alpha\right)}\right]-\frac{i}{\hbar}S\left[\phi_{+}=\varphi_{{\rm SP}}^{\left(\beta\right)}\right]}\,.\label{eq:4-110}
\end{eqnarray}
The sums over $\alpha$ and $\beta$ run independently over the full
set of stationary points. They pertain to, respectively, the $\phi_{+}$-
(or ``forward''-, or ``ket''-) sector and the $\phi_{-}$- (or
``backward''-, or ``bra''-) sector of Marinov's path integral. 

Recalling the discussion that led to Figure \ref{fig:3}, we see that
the index pairs $\left(\alpha,\beta\right)$ label a set of ruled
surfaces $\Sigma_{\left(\alpha,\beta\right)}$, which are defined
by classical data only. They are parametrized as in eq.~(\ref{eq:3-41})
by means of their ``directrices'' $\phi_{+}^{a}\left(t\right)$
and $\phi_{-}^{a}\left(t\right)$. In the case at hand, the latter
are the stationary point trajectories of type ``$\alpha$'' and
``$\beta$'', respectively:
\begin{eqnarray}
\chi_{\left(\alpha,\beta\right)}\left(t,s\right) & = & \frac{1}{2}\left(1-s\right)\varphi_{{\rm SP}}^{\left(\beta\right)}+\frac{1}{2}\left(1+s\right)\varphi_{{\rm SP}}^{\left(\alpha\right)}\,.\label{eq:4-111}
\end{eqnarray}
Furthermore, by virtue of eq.~(\ref{eq:3-SMS}), the exponentials
in (\ref{eq:4-110}) involve precisely the Marinov action $S_{{\rm M}}$
evaluated on those classical surfaces $\Sigma_{\left(\alpha,\beta\right)}$:
\begin{eqnarray}
 & \boxed{{\cal K}\left(0,q^{\prime\prime},T;0,q^{\prime},T_{0}\right)=\sum_{\Sigma_{\left(\alpha,\beta\right)}}e^{-2iS_{{\rm M}}\left[\chi_{\left(\alpha,\beta\right)}\right]}\,.}\label{eq:4-112}
\end{eqnarray}

The following points should be noted concerning this sum-over-surfaces
representation of the semiclassical probability density.

\noindent\textbf{(1) }The summation in (\ref{eq:4-112}) includes
also ``surfaces'' with $\alpha=\beta$, which actually degenerate
to a line as the boundaries $\partial\Sigma_{+}$ and $\partial\Sigma_{-}$
are on top of each other in this case. Since $S_{{\rm M}}\left[\chi_{\left(\alpha,\alpha\right)}\right]=0$,
they make a unit contribution to the sum (\ref{eq:4-112}).

\noindent\textbf{(2) }For $\alpha,\beta$ fixed, and $\alpha\neq\beta$,
the boundaries $\partial\Sigma_{\pm}$ are described by two different
solutions, $\varphi_{{\rm SP}}^{\left(\alpha\right)}$ and $\varphi_{{\rm SP}}^{\left(\beta\right)}$,
respectively. Using the $\left(\phi,\xi\right)$- rather than the
$\left(\phi_{+},\phi_{-}\right)$-language, a specific $\left(\alpha,\beta\right)$-pair
gives rise to the ``classical mechanics field''
\begin{eqnarray}
\phi_{\left(\alpha,\beta\right)}\left(t\right) & = & \frac{1}{2}\left(\varphi_{{\rm SP}}^{\left(\alpha\right)}\left(t\right)+\varphi_{{\rm SP}}^{\left(\beta\right)}\left(t\right)\right)\,,\label{eq:4-113}
\end{eqnarray}
and to the associated ``response field''
\begin{eqnarray}
\xi_{\left(\alpha,\beta\right)}\left(t\right) & = & \frac{1}{2\hbar}\left(\varphi_{{\rm SP}}^{\left(\alpha\right)}\left(t\right)-\varphi_{{\rm SP}}^{\left(\beta\right)}\left(t\right)\right)\,.\label{eq:4-114}
\end{eqnarray}
Together they parametrize $\Sigma_{\left(\alpha,\beta\right)}$ according
to
\begin{eqnarray}
\chi_{\left(\alpha,\beta\right)}\left(t,s\right) & = & \phi_{\left(\alpha,\beta\right)}\left(t\right)+s\hbar\,\xi_{\left(\alpha,\beta\right)}\left(t\right)\,.\label{eq:4-115added}
\end{eqnarray}
Identifying, as always, phase space points with vectors in $\mathbb{R}^{2N}$,
every pair of distinct saddle points gives rise to a non-zero response
field, which is given by the vectorial difference of $\varphi_{{\rm SP}}^{\left(\alpha\right)}\left(t\right)$
and $\varphi_{{\rm SP}}^{\left(\beta\right)}\left(t\right)$. 

Since $\varphi_{{\rm SP}}^{\left(\alpha\right)}\left(t\right)=\varphi_{{\rm SP}}^{\left(\beta\right)}\left(t\right)+2\hbar\xi_{\left(\alpha,\beta\right)}\left(t\right)$,
the vector $\xi_{\left(\alpha,\beta\right)}\left(t\right)$ connects
points on $\partial\Sigma_{-}$ to the corresponding points in $\partial\Sigma_{+}$,
and so it can be identified with one of the generators of $\Sigma_{\left(\alpha,\beta\right)}$.

Note also that $\xi_{\left(\alpha,\beta\right)}=-\xi_{\left(\beta,\alpha\right)}$
entails that $S_{{\rm M}}\left[\chi_{\left(\alpha,\beta\right)}\right]=-S_{{\rm M}}\left[\chi_{\left(\beta,\alpha\right)}\right]$.

\noindent\textbf{(3) }The boundary conditions (\ref{eq:4-104}) constrain
the $q$-components of $\varphi_{{\rm SP}}^{a}$. Therefore the projections
of the boundaries $\partial\Sigma^{\prime}\equiv\partial\Sigma_{\left(\alpha,\beta\right)}^{\prime}$
and $\partial\Sigma^{\prime\prime}\equiv\partial\Sigma_{\left(\alpha,\beta\right)}^{\prime\prime}$
onto the \emph{$q$}-subspace of \emph{${\cal M}$} are always degenerate:
they consist of one point only, viz.~$q^{\prime}$ and $q^{\prime\prime}$, respectively. 

\noindent\textbf{(4) }For every pair $\left(\alpha,\beta\right)$,
the $q$-space projection of the total boundary, $\partial\Sigma_{\left(\alpha,\beta\right)}$,
can be identified with \emph{a closed curve on configuration space},
henceforth denoted by $\Gamma_{\alpha,\beta}$. It comprises the projection
of $\partial\Sigma_{+}\cup\partial\Sigma_{-}$ with the terminal points
included. 

The curve $\Gamma_{\alpha,\beta}$ indeed closes since the respective
orientations are chosen such that $\varphi_{{\rm SP}}^{\left(\alpha\right)}\left(t\right)$
parametrizes $\partial\Sigma_{+}$, while $\varphi_{{\rm SP}}^{\left(\beta\right)}\left(t\right)$
is a parametrization of $-\partial\Sigma_{-}$. If $\alpha=\beta$,
the curve $\Gamma_{\alpha,\alpha}$ is backtracking: the return trajectory
from $q^{\prime\prime}$ back to $q^{\prime}$ lies atop the forward
trajectory from $q^{\prime}$ to $q^{\prime\prime}$.

\subsection{The double slit experiment}

The double slit experiment for electrons provides an instructive example
of the above setting and it is interesting also in its own right.
\begin{figure}
\captionsetup{justification=raggedright} 
\subfloat[]{\label{fig:sfig1}\includegraphics[scale=0.2]{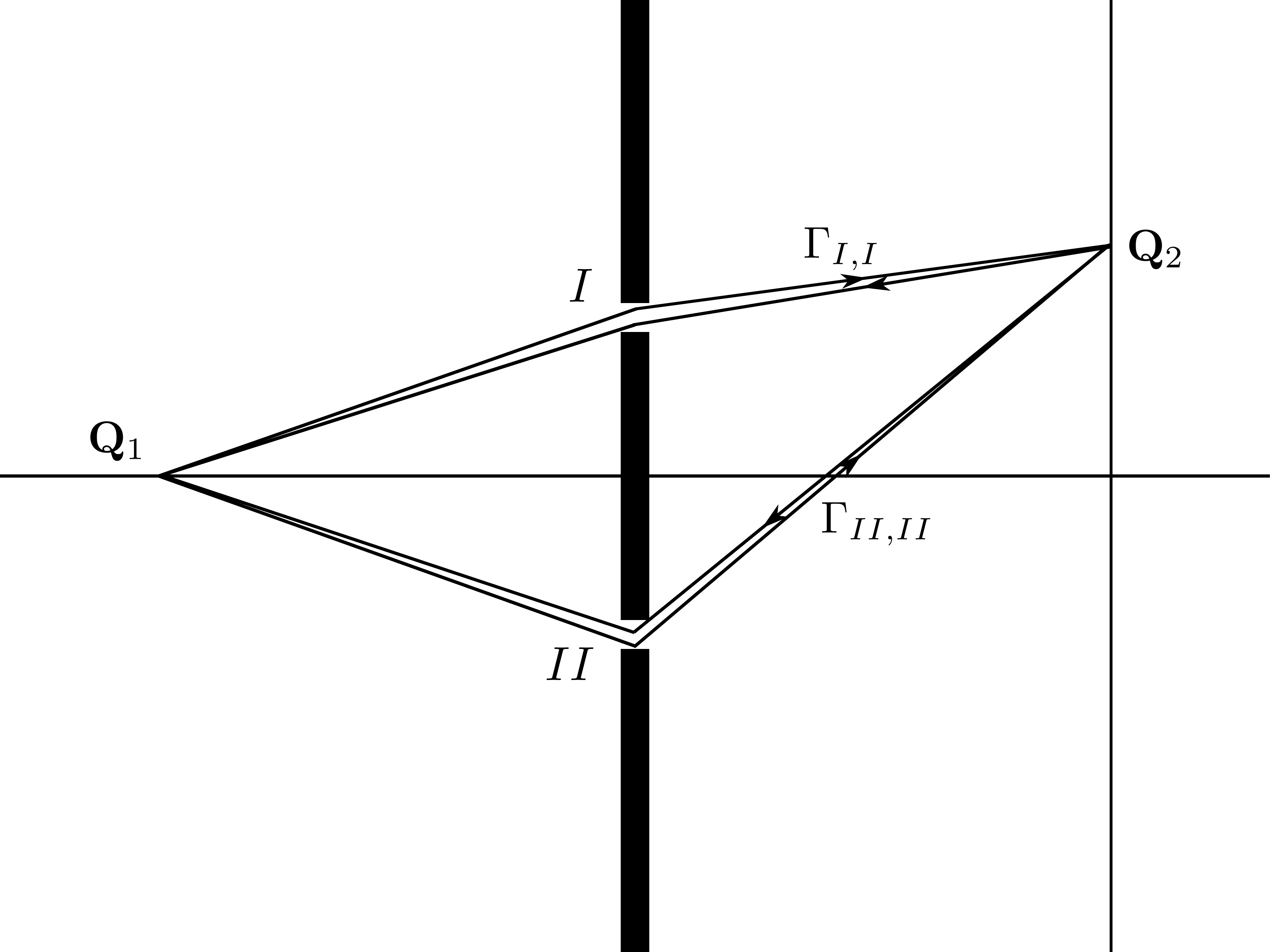}} 
\subfloat[]{\label{fig:sfig2}\includegraphics[scale=0.2]{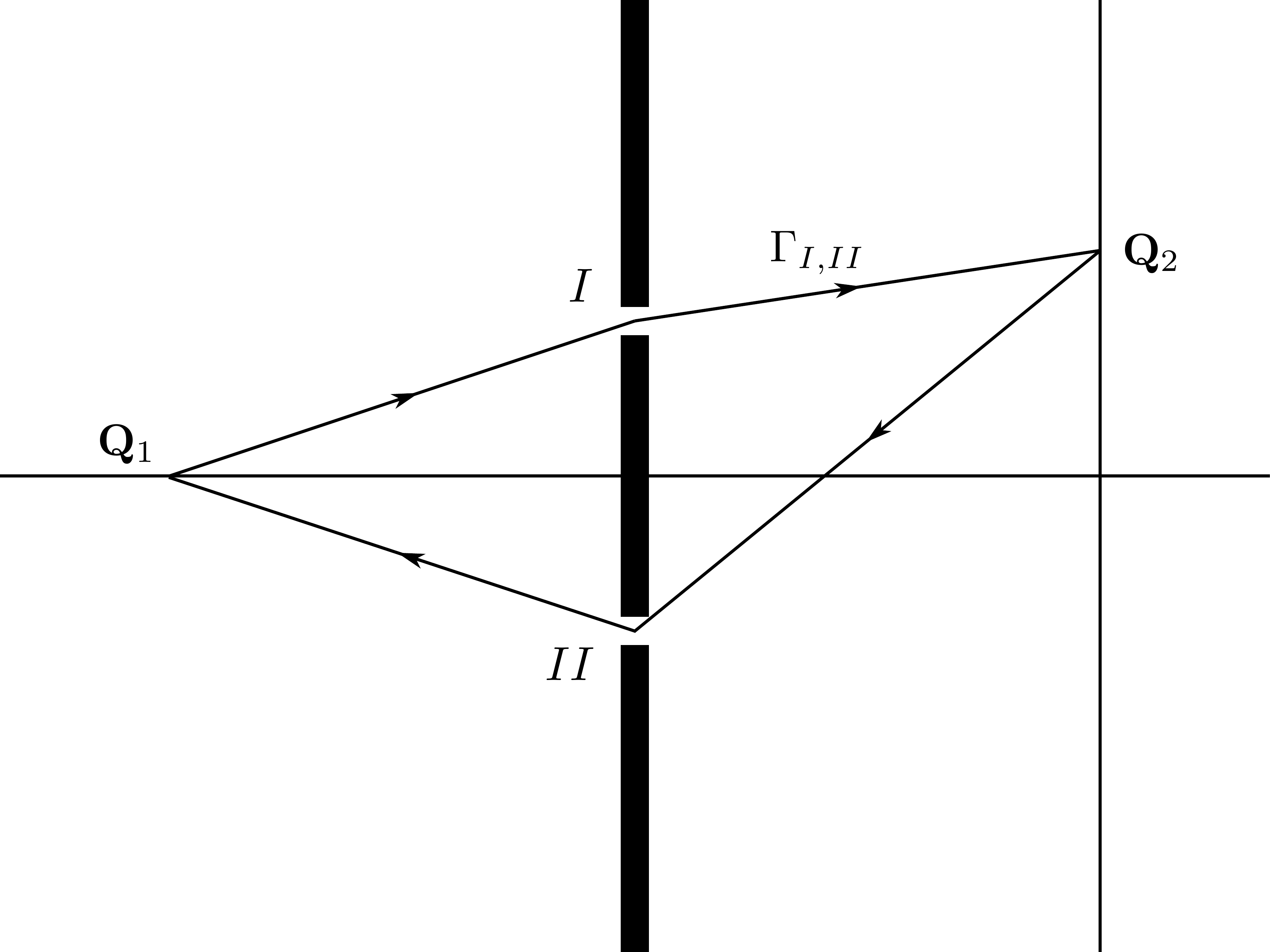}} 
\caption{\label{fig:4}
The closed curves $\Gamma_{\alpha,\beta}$ on the configuration space of the double slit experiment.
They are generated by the $q$-space projections of $\varphi_{\rm{SP}}^{(\alpha)}$ and $\varphi_{\rm{SP}}^{(\beta)}$, respectively.
Diagram (\ref{fig:sfig1}): The backtracking cases $\Gamma_{I,I}$ and $\Gamma_{II,II}$.  
Diagram (\ref{fig:sfig2}): The non-trivial cases $\Gamma_{I,II}=-\Gamma_{II,I}$ that are responsible for the interference structure.} 
\end{figure}

\noindent\textbf{(1) The setting.} 
As shown in Figure \ref{fig:4}, we consider
non-relativistic particles governed by the Hamiltonian\footnote{Here $N=3$, and as usual $3$-vectors are printed in boldface.}
$H=\frac{{\bf p}^{2}}{2m}+V\left({\bf q}\right)$ that are emitted
from an electron gun located at ${\bf Q}_{1}\widehat{=}\,q^{\prime}$,
then pass an obstacle having two slits, marked by $I$ and $II$,
and finally are detected on a screen or by a moveable detector at
the point ${\bf Q}_{2}\widehat{=}\,q^{\prime\prime}$. The only purpose
of the potential $V\left({\bf q}\right)$ is to physically model the
constraint imposed by the obstacle. Away from it the electrons propagate
freely.

\noindent\textbf{(2) Summing classical surfaces.} 
Let us employ the sum-over-surfaces representation
(\ref{eq:4-112}) in order to calculate the intensity of the electrons
hitting the screen. It is a function of $Q_{2}^{\perp}$, the vertical
distance of the point ${\bf Q}_{2}$ from the symmetry axis\footnote{The other argument, the time of flight $T-T_{0}$ is fixed by the
energy of the (by assumption, monoenergetic) electron beam.}:
\begin{eqnarray}
I\left(Q_{2}^{\perp}\right) & \propto & \left|K\left({\bf Q}_{2},T;{\bf Q}_{1},T_{0}\right)\right|^{2}\,=\,{\cal K}\left(0,{\bf Q}_{2},T;0,{\bf Q}_{1},T_{0}\right)\,.\label{eq:4-130}
\end{eqnarray}
In this example, there exist $2$ classical trajectories which can
play the role of the directrices, namely $\varphi_{{\rm SP}}^{\left(I\right)}\equiv\left({\bf p}_{{\rm SP}}^{\left(I\right)},{\bf x}_{{\rm SP}}^{\left(I\right)}\right)$
and $\varphi_{{\rm SP}}^{\left(II\right)}\equiv\left({\bf p}_{{\rm SP}}^{\left(II\right)},{\bf x}_{{\rm SP}}^{\left(II\right)}\right)$
whose ${\bf x}$-projections pass through the slits numbered $I$ and $II$, respectively.
We may assume that the momenta
\begin{eqnarray}
{\bf p}_{{\rm SP}}^{\left(I,II\right)} & = & m\,\dot{{\bf x}}_{{\rm SP}}^{\left(I,II\right)}\label{eq:4-131}
\end{eqnarray}
are time independent except for the sudden directional changes at the slits.

The two saddle points give rise to $4$ classical surfaces in the
phase space $\mathbb{R}^{3}\times\mathbb{R}^{3}$ which must be summed
over in eq.~(\ref{eq:4-112}): $\Sigma_{\left(I,I\right)}$, $\Sigma_{\left(II,II\right)}$,
$\Sigma_{\left(I,II\right)}$, and $\Sigma_{\left(II,I\right)}$.
The former two surfaces have backtracking configuration space projections,
implying $S_{{\rm M}}=0$, while the $S_{{\rm M}}$ values of the
latter two are equal up to a sign. Hence the sum over classical surfaces
in eq.~(\ref{eq:4-112}) leads to the result
\begin{eqnarray}
{\cal K}\left(0,{\bf Q}_{2},T;0,{\bf Q}_{1},T_{0}\right) & = & 1+1+e^{-2iS_{{\rm M}}\left[\chi_{\left(I,II\right)}\right]}+e^{+2iS_{{\rm M}}\left[\chi_{\left(I,II\right)}\right]}\nonumber \\
 & = & 4\cos^{2}\left(S_{{\rm M}}\left[\chi_{\left(I,II\right)}\right]\right)\,.\label{eq:4-132}
\end{eqnarray}
Thus, in order to interpret the physics 
behind the semiclassical sum over surfaces
it remains to understand
the role played by a single ruled surface, $\Sigma_{\left(I,II\right)}$.

\noindent\textbf{(3) The contributing surfaces and their projections.} 
In Figure \ref{fig:4} we sketch the ${\bf q}$-space
projections of the trajectories $\varphi_{{\rm SP}}^{\left(I,II\right)}$,
as well as the closed curves $\Gamma_{\alpha,\beta}$ they give rise
to. Figure \ref{fig:sfig1} depicts the backtracking curves $\Gamma_{I,I}$
and $\Gamma_{II,II}$ which ``know'' of one slit only, while Figure
\ref{fig:sfig2} shows $\Gamma_{I,II}=-\Gamma_{II,I}$ which involves
both slits and is at the heart of the interference effect therefore.

\begin{figure} 
\captionsetup{justification=raggedright}
\subfloat[]{\label{fig:5a}\includegraphics[scale=0.2]{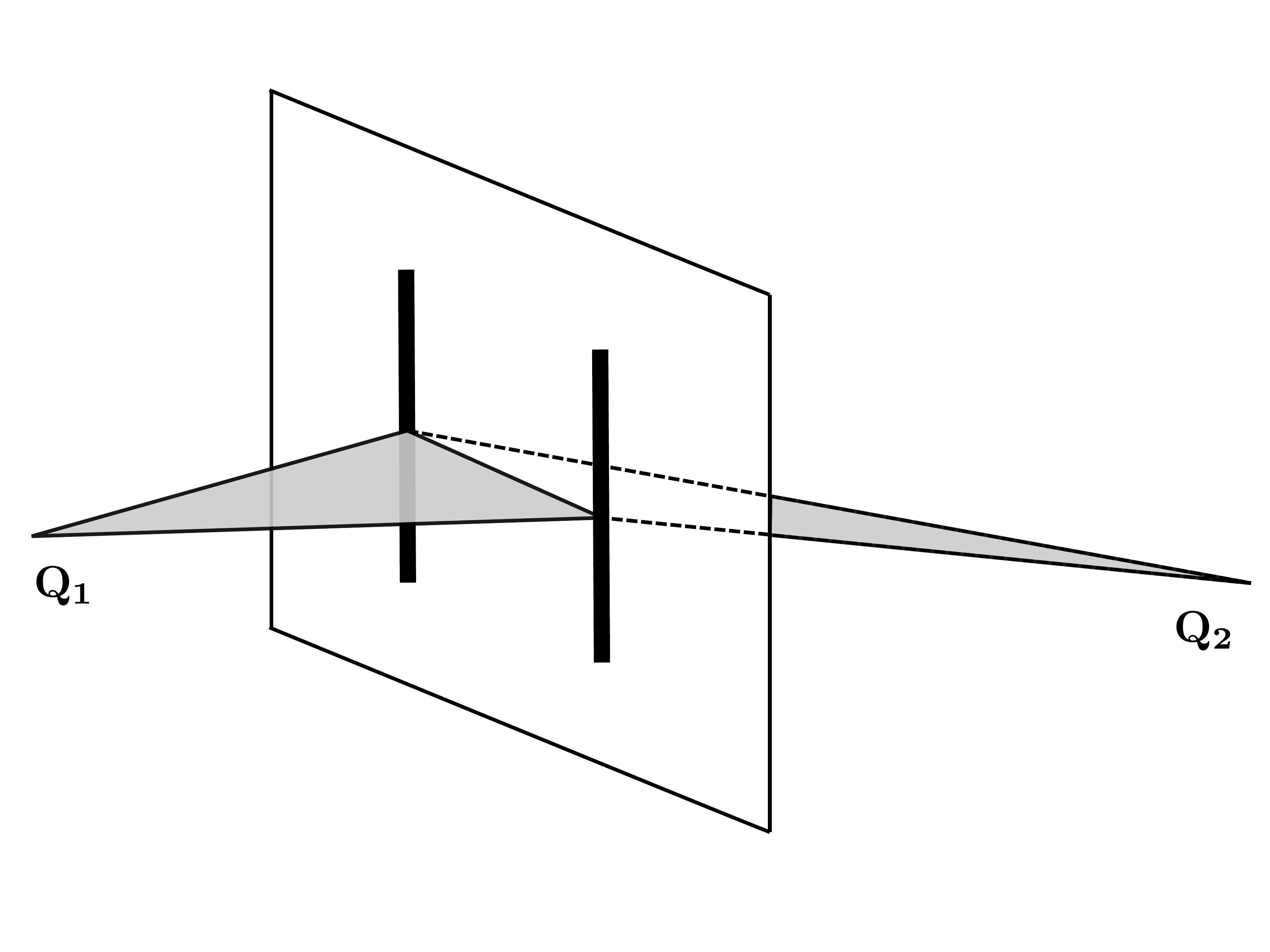}} 
\subfloat[]{\label{fig:5b}\includegraphics[scale=0.2]{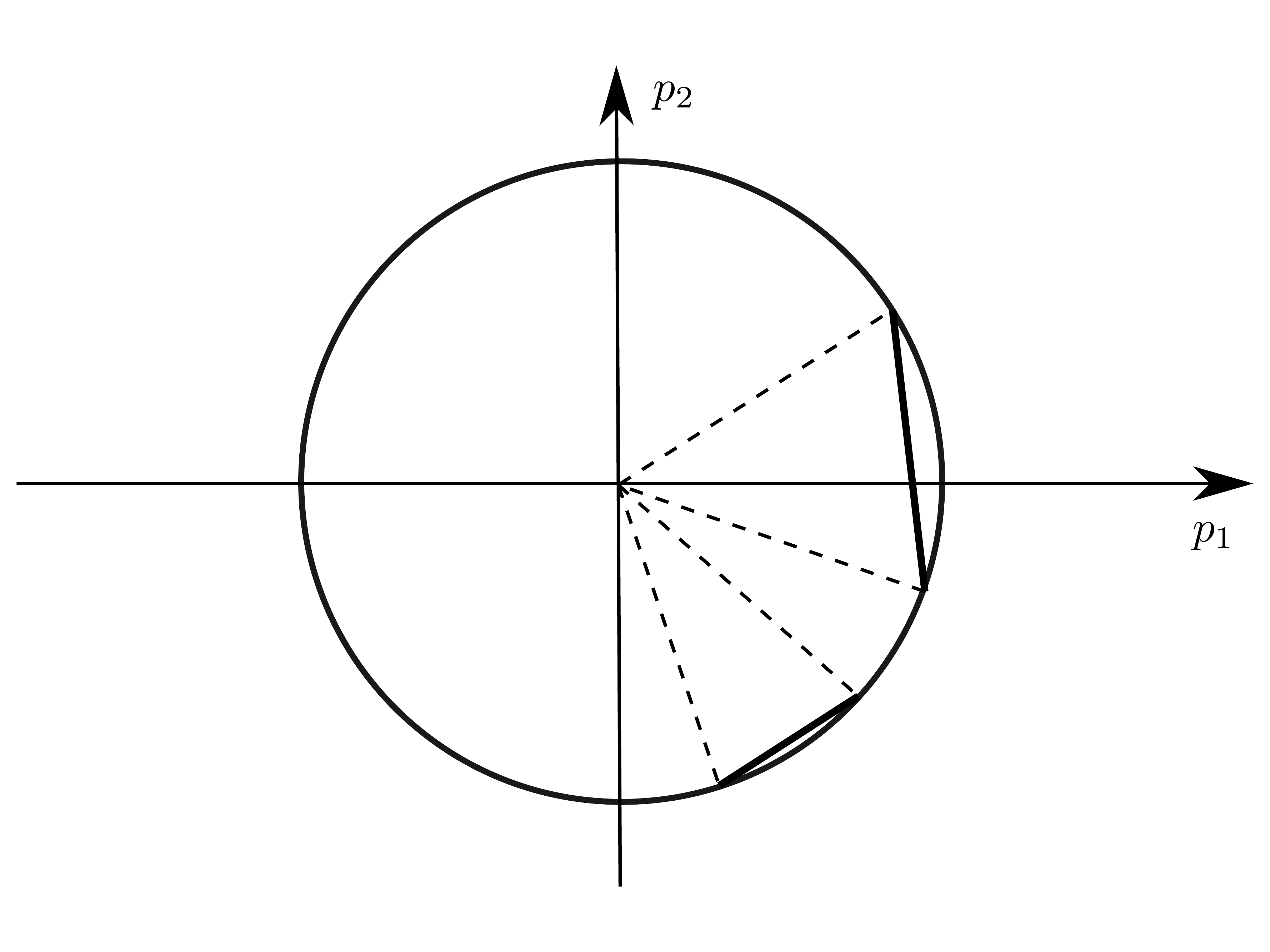}} 
\caption{\label{fig:5}
Diagram (\ref{fig:5a}): Projection of the surface $\Sigma_{(I,II)}$ from phase space onto configuration space. 
Diagram (\ref{fig:5b}): Projection of the surface $\Sigma_{(I,II)}$ onto the plane in momentum space perpendicular to the slits. The projection does not extend along the $p_3$ direction. It consists of two pieces of straight lines in the $p_1$-$p_2$ plane. Within the approximations considered, their endpoints lie all on the same circle with a radius given by the velocity of the free electrons. (All angles are heavily exaggerated.)} 
\end{figure}

In Figure \ref{fig:5} we try to visualize $\Sigma_{\left(I,II\right)}$,
actually a 2D surface in the $6$-dimensional phase space ${\cal M}=\mathbb{R}^{3}\times\mathbb{R}^{3}$,
by means of two projections. Figures \ref{fig:5a} and \ref{fig:5b}
schematically sketch the projections of the ``world-sheet'' onto the configuration
and momentum subspace, respectively. The former projection, $\Pi_{{\rm con}}\Sigma_{\left(I,II\right)}$,
consists of two planar triangles. They extend from the obstacle to
the electron source, and to the screen, respectively. The momentum
space projection on the other hand is one-dimensional, being comprised
of two straight line segments in the plane perpendicular to the slits.

\noindent\textbf{(3a)} 
Like the world sheet $\Sigma_{\left(I,II\right)}$ itself, the projections
are fully determined by classical information, namely the two trajectories
$\varphi_{{\rm SP}}^{\left(I,II\right)}$ passing through slits $I$
and $II$, respectively. The associated classical response field can
be visualized as the $6$-component difference vector
\begin{eqnarray}
\varphi_{{\rm SP}}^{\left(I\right)}\left(t\right)-\varphi_{{\rm SP}}^{\left(II\right)}\left(t\right) & \equiv & 2\hbar\,\xi_{\left(I,II\right)} \left(t\right)\,.\label{eq:diff-V}
\end{eqnarray}
Loosely speaking, at any given time, $\xi_{\left(I,II\right)}\left(t\right)$
describes a rigid, straight string in phase space which is spanned
between the only two points at which a \emph{classical} particle could
be found at that time, $\varphi_{{\rm SP}}^{\left(I\right)}\left(t\right)$
and $\varphi_{{\rm SP}}^{\left(II\right)}\left(t\right)$.

In this picture, the phase space variables employed in classical mechanics
amounts to the ``center of mass'' of those two locations, situated
precisely in the middle of the string:
\begin{eqnarray}
\frac{1}{2}\left[\varphi_{{\rm SP}}^{\left(I\right)}\left(t\right)+\varphi_{{\rm SP}}^{\left(II\right)}\left(t\right)\right] & \equiv & \phi_{\left(I,II\right)}\left(t\right)\,.\label{eq:CM-V}
\end{eqnarray}
 
\noindent\textbf{(3b)} 
It is rather striking how the beam of electrons interacts
with the obstacle in the string picture: $\Pi_{{\rm con}}\Sigma_{\left(I,II\right)}$,
the ${\bf x}$-projection of the world-sheet swept out by the propagating string
is by no means forced to terminate at the obstacle,
as one might have expected.
Rather, it passes
through the obstacle and merely changes its direction there.

Indeed, the projected string world sheet feels the potential $V\left({\bf q}\right)$
only along its boundaries formed by $\varphi_{{\rm SP}}^{\left(I\right)}\left(t\right)$
and $\varphi_{{\rm SP}}^{\left(II\right)}\left(t\right)$. 
The latter are
perfectly ordinary classical trajectories, which are guided by the
potential through the respective slits. The \emph{inner} points of
the moving string, on the other hand, follow a curve
\begin{eqnarray}
t & \mapsto & \chi_{\left(I,II\right)}\left(s,t\right)=\frac{1}{2}\left(1-s\right)\varphi_{{\rm SP}}^{\left(I\right)}\left(t\right)+\frac{1}{2}\left(1+s\right)\varphi_{{\rm SP}}^{\left(II\right)}\left(t\right)\label{eq:interpol}
\end{eqnarray}
for some fixed value of $s$ different from $\pm1$. This curve is
{\it not} a classical solution, however, and so there is no general reason 
that would
prevent it from passing right through the obstacle. Nevertheless,
since (\ref{eq:interpol}) is a ``linear interpolation'' of two
classical trajectories, also the inner points change their direction
of motion at the obstacle.

\noindent\textbf{(3c)} 
It is particularly remarkable that also the symmetrized variable on which conventional
classical mechanics is based, 
$\phi\left(t\right)=\frac{1}{2}\left[\phi_{+}\left(t\right)+\phi_{-}\left(t\right)\right]$,
parametrizes such an inner point of the string, namely its center:
$\chi_{\left(I,II\right)}\left(s,t\right)\Bigr|_{s=0}=\phi_{\left(I,II\right)}\left(t\right)$.

For precisely this reason it is impossible to meaningfully interpret
the outcome of the double slit experiment within strictly classical
mechanics: On the one side, the limit $\hbar\rightarrow0$ of quantum
mechanics {\`a} la Marinov furnishes the resulting classical theory
with dynamical variables of the correct type $\phi\equiv\frac{1}{2}\left(\phi_{+}+\phi_{-}\right)$,
but on the other side, the concrete function
$t\mapsto\phi\left(t\right)=\chi_{(I,II)}\left(0,t\right)$
which it delivers in the classical limit does
\emph{not} satisfy the classical equation of motion.

Clearly, from the quantum a.k.a.~string perspective, there is nothing
wrong about this apparent paradox: Three different curves play a role
here, namely $\varphi_{{\rm SP}}^{\left(I\right)}$, $\varphi_{{\rm SP}}^{\left(II\right)}$,
and $\phi_{\left(I,II\right)}$, and generically they may differ by
potentially large contributions proportional to the response field
$\xi_{\left(I,II\right)}$.

\noindent\textbf{(4) The interference pattern.}
Given the saddle point worldsheets, it is straightforward
now to complete the calculation, to write down the two directrices
$\varphi_{{\rm SP}}^{\left(I,II\right)}$ and the parametrization
of $\Sigma_{\left(I,II\right)}$, and to evaluate the world sheet
action $S_{{\rm M}}$ on this ruled surface. Under the assumption
that the distance between the slits, $a$, is much smaller than the
distances from the obstacle to both the screen and the electron gun,
we recover exactly the standard result from our eq.~(\ref{eq:4-132}):
\begin{eqnarray}
I\left(Q_{2}^{\perp}\right) & \propto & 4\cos^{2}\left(\frac{mva}{2\hbar\ell}Q_{2}^{\perp}\right)\,.\label{eq:4-139}
\end{eqnarray}
Here $v$ denotes the electron velocity, $\ell$ the distance obstacle-screen,
and $Q_{2}^{\perp}$ is the above coordinate on the screen in the
direction perpendicular to the slits. The modulation of the intensisty
(\ref{eq:4-139}) displays precisely the expected interference pattern.
(Details of the calculation can be found in \cite{Felsager:1981iy,Gozzi:2020wef}.)

\noindent\textbf{(5) Tuning the size of the worldsheet.}
A typical feature of this calculation is that the size of the
strings relevant in the semiclassical regime, and thus the extension of
their worldsheet in phase space, can be controlled by an externally accessible
parameter, the distance of the slits, $a$, being a example.

In fact, from eq.~(\ref{eq:diff-V}) we can infer that at the moment
when 
$\varphi_{{\rm SP}}^{\left(I\right)}$ and $\varphi_{{\rm SP}}^{\left(II\right)}$
pass through the slits, the response field controlling the size of
the projected world sheet is 
forced to acquire a ${\bf q}$-component $\left(2\hbar\right) \xi^q$ as large as
the distance $a$.
Upon extracting the ${\bf q}$-components of eq.~(\ref{eq:diff-V}), and
forming the norm of the resulting 3-vector, we obtain indeed $\left(2\hbar\right)\left\Vert \xi_{\left(I,II\right)}^{q}\left(t\right)\right\Vert =a$.
Thus, even in the semiclassical regime, the magnitude of the relevant response
fields and the size of the resulting world sheets can be tuned to assume macroscopic
dimensions.

\section{Summary and conclusions} \label{sec:conclusions}

In this work we derived a novel representation for the evolution kernel of Wigner functions in terms of random surfaces. 
It turns out that the surfaces involved in this representation are not generic ones but the so-called ruled surfaces. 
The introduction of the ruled surfaces (\ref{eq:3-35}) allowed us to embed the $4N$ variables appearing in Marinov integral into the $2N$ surfaces $\chi^{a}$. 
Remarkably, the action that characterizes the new path integral, 
eq.~(\ref{eq:3-85}), can be given a geometric interpretation in terms of these surfaces. In this sense, one can look at our representation as an alternative to the formulation of quantization via branes \cite{Witten:2010zr,Gukov:2008ve}.

In section \ref{sec:sum_classical_surf} we introduced the semiclassical approximation and studied the double slit experiment in this setting. The picture that emerges is rather striking: the size of the ``string'' is related to the distance between the two slits and, in a loose sense, it carries memory of the non-local effects of quantum mechanics. This example hints towards the fact that the immaterial string gives rise to a more local picture of quantum correlations over large distances that otherwise may appear non-local. In a similar configuration as the one used in section 4, in our previous work \cite{Gozzi:2020wef} we studied the Aharonov-Bohm effect. In this latter case the string world sheet covers the flux tube, thus reinforcing a more local picture of quantum processes.

This works opens the possibility of investigating whether novel approximation schemes may be constructed based on the random surface representation of the evolution kernel. It would also be interesting to study if it is possible to further generalize our surface representation in terms of more general objects.


\newpage{}

\appendix

\section{Path integral on a time lattice \label{sec:Path-integral-on-time-lattice}}

We recall the discretized form of the quantum Hamiltonian path integral
\begin{eqnarray*}
K\left(x_{F},t;x_{I},0\right) & = & \lim_{n\rightarrow\infty}\int\prod_{j=1}^{n-1}dx_{j}\prod_{j=1}^{n}dp_{j}\left(\frac{1}{2\pi\hbar}\right)^{n}e^{\sum_{j=1}^{n}i\frac{1}{\hbar}\left[p_{j}\frac{x_{j}-x_{j-1}}{\varepsilon}-H\left(p_{j},\bar{x}_{j-1}\right)\right]\varepsilon}\\
 & \equiv & \int_{x\left(0\right)=x_{I}}^{x\left(t\right)=x_{F}}{\cal D}x{\cal D}p\,e^{i\frac{1}{\hbar}\left[p\dot{x}-H\right]}\,,
\end{eqnarray*}
where $x_{0}=x_{I}$, $x_{n}=x_{F}$, and $\bar{x}_{j}\equiv\left(1-\alpha\right)x_{j-1}+\alpha x_{j}$
defines a discretization prescription for $\alpha\in\left[0,1\right]$.
From now on we will work with the pre-point discretization convention
$\bar{x}_{j-1}=x_{j-1}$. Now consider
\begin{eqnarray*}
K_{{\rm M}}\left(q,p,t;q_{0},p_{0},0\right) & = & \frac{1}{2\pi\hbar}\int ds\,ds_0 
\,e^{\frac{i}{\hbar}\left(s_{0}p_{0}-sp\right)}\underbrace{K\left(q+\frac{s}{2},t;q_{0}+\frac{s_{0}}{2},0\right)K\left(q-\frac{s}{2},t;q_{0}-\frac{s_{0}}{2},0\right)^{\dagger}}_{\equiv{\cal K}}\,.
\end{eqnarray*}

It is convenient to introduce the following variables:
\begin{eqnarray*}
x_{s,j} & \equiv & \frac{x_{+,j}+x_{-,j}}{2}\\
x_{a,j} & \equiv & x_{+,j}-x_{-,j}\\
x_{a,0} & \equiv & s_{0}\\
x_{a,n} & \equiv & s\,.
\end{eqnarray*}
Similar combinations are introduced for the momenta $p_{\pm}$. One
has
\begin{eqnarray*}
{\cal K} & = & \int\prod_{j=1}^{n-1}dx_{+,j}\prod_{j=1}^{n}dp_{+,j}\left(\frac{1}{2\pi\hbar}\right)^{n}\exp\left\{ \sum_{j=1}^{n}\frac{i}{\hbar}\left[p_{+,j}\frac{x_{+,j}-x_{+,j-1}}{\varepsilon}-H\left(p_{+,j},x_{+,j-1}\right)\right]\varepsilon\right\} \\
 &  & \int\prod_{j=1}^{n-1}dx_{-,j}\prod_{j=1}^{n}dp_{-,j}\left(\frac{1}{2\pi\hbar}\right)^{n}\exp\left\{ -\sum_{j=1}^{n}\frac{i}{\hbar}\left[p_{-,j}\frac{x_{-,j}-x_{-,j-1}}{\varepsilon}-H\left(p_{-,j},x_{-,j-1}\right)\right]\varepsilon\right\}
\end{eqnarray*}
with $x_{+,0}=q_{0}+\frac{s_{0}}{2}$, 
$x_{-,0}=q_{0}-\frac{s_{0}}{2}$,
$x_{+,n}=q+\frac{s}{2}$,
and $x_{-,n}=q-\frac{s}{2}$.
By going to the symmetric and anti-symmetric variables 
and noticing that
$x_{s,0}=q_{0}$,
$x_{a,0}=s_{0}$,
$x_{s,n}=q$, and
$x_{a,n}=s$
one can rewrite
\begin{eqnarray*}
K_{{\rm M}}\left(q,p,t;q_{0},p_{0},0\right) & = & \frac{1}{2\pi\hbar}\int ds\,ds_{0}\,e^{\frac{i}{\hbar}\left(s_{0}p_{0}-sp\right)}{\cal K}\left(q,s,t;q_{0},s_{0},0\right)\\
 & = & \frac{1}{2\pi\hbar}\int dx_{a,n}\,dx_{a,0}\,e^{\frac{i}{\hbar}\left(x_{a,0}p_{0}-x_{a,n}p\right)}\\
 &  & \int\prod_{j=1}^{n-1}dx_{s,j}\prod_{j=1}^{n}dp_{s,j}\int\prod_{j=1}^{n-1}dx_{a,j}\prod_{j=1}^{n}dp_{a,j}\left(\frac{1}{2\pi\hbar}\right)^{2n}\\
 &  & \exp\left\{ \sum_{j=1}^{n}\frac{i}{\hbar}\left[p_{+,j}\frac{x_{+,j}-x_{+,j-1}}{\varepsilon}-H\left(p_{+,j},x_{+,j-1}\right)\right]\varepsilon\right\} \\
 &  & \exp\left\{ -\sum_{j=1}^{n}\frac{i}{\hbar}\left[p_{-,j}\frac{x_{-,j}-x_{-,j-1}}{\varepsilon}-H\left(p_{-,j},x_{-,j-1}\right)\right]\varepsilon\right\} \,,
\end{eqnarray*}
where the integration is now over the symmetric and antisymmetric variables,
the Fourier transform is written in terms of $x_{a,0}$ and  $x_{a,n}$,
and
the variables $x_{\pm}$ and $p_\pm$ are now functions of the symmetric and antisymmetric ones.

In order to bring the path integral into the desired form 
let us rewrite the terms $p\dot{q}$ in terms of $x_{s,a}$ and $p_{s,a}$
(including also the terms from the Fourier transform):
\begin{eqnarray*}
\left(x_{a,0}p_{0}-x_{a,n}p\right)+\varepsilon\sum_{j=1}^{n}p_{+,j}\frac{x_{+,j}-x_{+,j-1}}{\varepsilon}-\varepsilon\sum_{j=1}^{n}p_{-,j}\frac{x_{-,j}-x_{-,j-1}}{\varepsilon} & =\\
\left(x_{a,0}p_{0}-x_{a,n}p\right)+\sum_{j=1}^{n}\left(p_{s,j}+\frac{p_{a,j}}{2}\right)\left(x_{s,j}+\frac{x_{a,j}}{2}-x_{s,j-1}-\frac{x_{a,j-1}}{2}\right)\\
-\sum_{j=1}^{n}\left(p_{s,j}-\frac{p_{a,j}}{2}\right)\left(x_{s,j}-\frac{x_{a,j}}{2}-x_{s,j-1}+\frac{x_{a,j-1}}{2}\right) & =\\
\left(x_{a,0}p_{0}-x_{a,n}p\right)+\sum_{j=1}^{n}\frac{p_{a,j}}{2}2\left(x_{s,j}-x_{s,j-1}\right)\\
+\sum_{j=1}^{n}p_{s,j}\left(x_{a,j}-x_{a,j-1}\right)\,.
\end{eqnarray*}
It is convenient to rewrite the sum involving the $x_{a}$ as follows
\begin{eqnarray*}
\left(x_{a,0}p_{0}-x_{a,n}p\right)+\sum_{j=1}^{n}p_{s,j}\left(x_{a,j}-x_{a,j-1}\right) & =\\
-x_{a,n}\left(p-p_{s,n}\right)-\sum_{j=1}^{n-1}x_{a,j}\left(p_{s,j+1}-p_{s,j}\right)-x_{a,0}\left(p_{s,1}-p_{0}\right) & =\\
\mbox{denote }p_{0}\equiv p_{s,0}\\
-x_{a,n}\left(p-p_{s,n}\right)-\sum_{j=0}^{n-1}x_{a,j}\left(p_{s,j+1}-p_{s,j}\right)\,.
\end{eqnarray*}

Therefore the full discretized kernel can be written as follows:
\begin{eqnarray*}
K_{{\rm M}} & = & \frac{1}{2\pi\hbar}\int dx_{a,n}\,dx_{a,0}\int\prod_{j=1}^{n-1}dx_{s,j}\prod_{j=1}^{n}dp_{s,j}\int\prod_{j=1}^{n-1}dx_{a,j}\prod_{j=1}^{n}dp_{a,j}\left(\frac{1}{2\pi\hbar}\right)^{2n}\\
 &  & \exp\left\{ \frac{i}{\hbar}\left[-x_{a,n}\left(\frac{p-p_{s,n}}{\varepsilon}\right)-\sum_{j=0}^{n-1}x_{a,j}\left(\frac{p_{s,j+1}-p_{s,j}}{\varepsilon}\right)+\sum_{j=1}^{n}p_{a,j}\left(\frac{x_{s,j}-x_{s,j-1}}{\varepsilon}\right)\right]\varepsilon\right\} \\
 &  & \exp\left\{ \sum_{j=1}^{n}\frac{i}{\hbar}\left[H\left(p_{s,j}-\frac{p_{a,j}}{2},x_{s,j-1}-\frac{x_{a,j-1}}{2}\right)-H\left(p_{s,j}+\frac{p_{a,j}}{2},x_{s,j-1}+\frac{x_{a,j-1}}{2}\right)\right]\varepsilon\right\} \,.
\end{eqnarray*}
It is now crucial to observe that in the discretized Hamiltonian of
the third line above there is no dependence on $x_{a,n}$. Therefore,
we can carry out the $x_{a,n}$-integration straightforwardly:
\begin{eqnarray*}
K_{{\rm M}} & = & \cancel{\frac{1}{2\pi\hbar}}\int dx_{a,0}\int\prod_{j=1}^{n-1}dx_{s,j}\prod_{j=1}^{n}dp_{s,j}\int\prod_{j=1}^{n-1}dx_{a,j}\prod_{j=1}^{n}dp_{a,j}\left(\frac{1}{2\pi\hbar}\right)^{2n}\\
 &  & \cancel{2\pi\hbar}\delta\left(p-p_{s,n}\right)\exp\left\{ \frac{i}{\hbar}\left[-\sum_{j=0}^{n-1}x_{a,j}\left(\frac{p_{s,j+1}-p_{s,j}}{\varepsilon}\right)+\sum_{j=1}^{n}p_{a,j}\left(\frac{x_{s,j}-x_{s,j-1}}{\varepsilon}\right)\right]\varepsilon\right\} \\
 &  & \exp\left\{ \sum_{j=1}^{n}\frac{i}{\hbar}\left[H\left(p_{s,j}-\frac{p_{a,j}}{2},x_{s,j-1}-\frac{x_{a,j-1}}{2}\right)-H\left(p_{s,j}+\frac{p_{a,j}}{2},x_{s,j-1}+\frac{x_{a,j-1}}{2}\right)\right]\varepsilon\right\} \\
 &  & \mbox{integrate over }p_{s,n}\\
 & = & \int dx_{a,0}\int\prod_{j=1}^{n-1}dx_{s,j}\prod_{j=1}^{n-1}dp_{s,j}\int\prod_{j=1}^{n-1}dx_{a,j}\prod_{j=1}^{n}dp_{a,j}\left(\frac{1}{2\pi\hbar}\right)^{2n}\\
 &  & \exp\left\{ \frac{i}{\hbar}\left[-\sum_{j=0}^{n-1}x_{a,j}\left(\frac{p_{s,j+1}-p_{s,j}}{\varepsilon}\right)+\sum_{j=1}^{n}p_{a,j}\left(\frac{x_{s,j}-x_{s,j-1}}{\varepsilon}\right)\right]\varepsilon\right\} \\
 &  & \exp\left\{ \sum_{j=1}^{n}\frac{i}{\hbar}\left[H\left(p_{s,j}-\frac{p_{a,j}}{2},x_{s,j-1}-\frac{x_{a,j-1}}{2}\right)-H\left(p_{s,j}+\frac{p_{a,j}}{2},x_{s,j-1}+\frac{x_{a,j-1}}{2}\right)\right]\varepsilon\right\} \,,
\end{eqnarray*}
with $p_{s,n}=p$, $p_{s,0}=p_{0}$, $x_{s,n}=q$, and $x_{s,0}=q_{0}$.
We can relabel the $x_{a,j}\rightarrow x_{a,j+1}$ so that finally
one can rewrite
\begin{eqnarray*}
K_{{\rm M}} & = & \int\prod_{j=1}^{n-1}dx_{s,j}\prod_{j=1}^{n-1}dp_{s,j}\int\prod_{j=1}^{n}dx_{a,j}\prod_{j=1}^{n}dp_{a,j}\left(\frac{1}{2\pi\hbar}\right)^{2n}\\
 &  & \exp\left\{ \frac{i}{\hbar}\left[-\sum_{j=1}^{n}x_{a,j}\left(\frac{p_{s,j}-p_{s,j-1}}{\varepsilon}\right)+\sum_{j=1}^{n}p_{a,j}\left(\frac{x_{s,j}-x_{s,j-1}}{\varepsilon}\right)\right]\varepsilon\right\} \\
 &  & \exp\left\{ \sum_{j=1}^{n}\frac{i}{\hbar}\left[H\left(p_{s,j}-\frac{p_{a,j}}{2},x_{s,j-1}-\frac{x_{a,j}}{2}\right)-H\left(p_{s,j}+\frac{p_{a,j}}{2},x_{s,j-1}+\frac{x_{a,j}}{2}\right)\right]\varepsilon\right\} \,,
\end{eqnarray*}
which is the discretized path integral in the desired form.
By using a continuum notation one then has
\begin{eqnarray*}
K_{{\rm M}} & = & \int{\cal D}x_{s}\int{\cal D}p_{s}\int{\cal D}x_{a}\int{\cal D}p_{a}\\
 &  & \exp\left\{ \frac{i}{\hbar}\left[-\int x_{a}\dot{p}_{s}+\int p_{a}\dot{x}_{s}\right]+\frac{i}{\hbar}\int\left[H\left(p_{s}-\frac{p_{a}}{2},x_{s}-\frac{x_{a}}{2}\right)-H\left(p_{s}+\frac{p_{a}}{2},x_{s}+\frac{x_{a}}{2}\right)\right]\right\} \\
 & = & \int{\cal D}x_{s}\int{\cal D}p_{s}\int{\cal D}x_{a}\int{\cal D}p_{a}\\
 &  & \exp\left\{ \frac{i}{\hbar}\left[\left(-=\omega_{qp}\right)\int x_{a}\dot{\phi}^{p}+\left(+=\omega_{pq}\right)\int p_{a}\dot{\phi}^{q}\right]\right\} \\
 &  & \exp\left\{ -\frac{i}{\hbar}\int\left[H\left(p_{s}+\frac{p_{a}}{2},x_{s}+\frac{x_{a}}{2}\right)-H\left(p_{s}-\frac{p_{a}}{2},x_{s}-\frac{x_{a}}{2}\right)\right]\right\} \,.
\end{eqnarray*}
By introducing $\phi^{a}\equiv\left(p_{s},x_{s}\right)$ and $\check{\xi}^{a}\equiv\left(p_{a},x_{a}\right)$
one obtains
\begin{eqnarray*}
K_{{\rm M}} & = & \int{\cal D}x_{s}\int{\cal D}p_{s}\int{\cal D}x_{a}\int{\cal D}p_{a}\exp\left\{ \frac{i}{\hbar}\int\check{\xi}^{a}\omega_{ab}\dot{\phi}^{b}=-\frac{i}{\hbar}\int\dot{\phi}^{a}\omega_{ab}\check{\xi}^{b}\right\} \\
 &  & \exp\left\{ -\frac{i}{\hbar}\int\left[H\left(\left\{ p_{s}+\frac{p_{a}}{2},x_{s}+\frac{x_{a}}{2}\right\} =\phi^{a}+\frac{1}{2}\check{\xi}^{a}\right)-H\left(\phi^{a}-\frac{1}{2}\check{\xi}^{a}\right)\right]\right\} \\
 & = & \int{\cal D}x_{s}\int{\cal D}p_{s}\int{\cal D}x_{a}\int{\cal D}p_{a}\\
 &  & \exp\left\{ -\frac{i}{\hbar}\int\left[\dot{\phi}^{a}\omega_{ab}\check{\xi}^{b}+\left(H\left(\phi^{a}+\frac{1}{2}\check{\xi}^{a}\right)-H\left(\phi^{a}-\frac{1}{2}\check{\xi}^{a}\right)\right)\right]\right\} \\
 & = & \int{\cal D}x_{s}\int{\cal D}p_{s}\int{\cal D}x_{a}\int{\cal D}p_{a}\\
 &  & \exp\left\{ -\frac{i}{\hbar}\int\left[\dot{\phi}^{a}\omega_{ab}\check{\xi}^{b}-\left(H\left(\phi^{a}-\frac{1}{2}\check{\xi}^{a}\right)-H\left(\phi^{a}+\frac{1}{2}\check{\xi}^{a}\right)\right)\right]\right\} \,.
\end{eqnarray*}
By further defining $\check{\xi}^{a}=2\hbar\xi^{a}$ one obtains
\begin{eqnarray*}
K_{{\rm M}} & = & \int{\cal D}\phi^{a}\int{\cal D}\xi^{a}\exp\left\{ -2i\int\left[\dot{\phi}^{a}\omega_{ab}\xi^{b}-\frac{1}{2\hbar}\left(H\left(\phi^{a}-\hbar\xi^{a}\right)-H\left(\phi^{a}+\hbar\xi^{a}\right)\right)\right]\right\} \,.
\end{eqnarray*}

\bibliography{paper}

\end{document}